\documentclass[preprint,12pt,authoryear]{elsarticle}

\usepackage{amsthm}
\usepackage{xcolor}
\usepackage{mathrsfs}
\usepackage{lipsum}
\usepackage{amsfonts}
\usepackage{algorithmic}
\usepackage{amssymb}
\usepackage{amsmath}
\usepackage{graphicx}
\usepackage{epstopdf}
\usepackage{subfig}
\usepackage{mathtools}
\usepackage{tabularx}
\usepackage{algorithm2e}
\usepackage{wrapfig}
\usepackage{hyperref}
\usepackage{cleveref}
\usepackage{bm}
\usepackage{amsopn}
\usepackage{sidecap}
\usepackage{placeins}
\usepackage{float}

\SetKwComment{Comment}{/* }{ */}

\DeclareMathOperator{\Tr}{tr}
\ifpdf
\DeclareGraphicsExtensions{.eps,.pdf,.png,.jpg}
\else
\DeclareGraphicsExtensions{.eps}
\fi
\newcommand{\RN}[1]{%
  \textup{\uppercase\expandafter{\romannumeral#1}}%
}
\newcommand{\rn}[1]{%
  \textup{\expandafter{\romannumeral#1}}%
}
\usepackage{tikz}
\usetikzlibrary{arrows.meta}
\usetikzlibrary{decorations.pathreplacing}

\newtheorem*{lemma*}{Lemma}
\newtheorem*{corollary*}{Corollary}
\newtheorem*{remark*}{Remark}

\theoremstyle{definition}

\journal{Journal of Theoretical Biology}

\begin{document}

\begin{frontmatter}



\title{Continuum modeling of fluidic and elastic flow during growth-driven wound closure in partial-EMT cell monolayers}

\author[add1,add7]{Chaozhen Wei}
\author[add3,add7]{Han Jiang}
\author[add1]{Yifan Gu}
\author[add1]{Nonthakorn Olaranont}
\author[add4]{Pengbo Wang}
\author[add4]{Qi Wen}
\author[add3]{Yubing Sun\corref{cor1}}
\ead{ybsun@umass.edu}
\author[add1]{Min Wu\corref{cor1}}
\ead{englier@gmail.com}


\address[add1]{Department of Mathematical Sciences, Worcester Polytechnic Institute}
\address[add3]{Department of Mechanical and Industrial Engineering, University of Massachusetts, Amherst}
\address[add4]{Department of Physics, Worcester Polytechnic Institute}
\address[add7]{The authors contribute equally.}

\cortext[cor1]{Corresponding authors.}


\begin{abstract}
Large-scale circular gap closure occurs over a time scale on which cell growth and proliferation become important. Growth is the main driver of the closing process, while cell dynamics such as elongation and intercalation reflect elastic and fluidic contributions to tissue deformation. We develop a novel fluidized growth-elasticity framework as a nonlinear analogue of a Maxwell fluid with growth. The framework decomposes the experimentally observable strain rate into the additive sum of the growth, elastic, and fluidic strain rates, thus enabling the separate quantification of these contributions from tissue kinematics and allowing the roles of tissue elasticity and fluidity (the inverse of viscosity) to be characterized. We apply the model to large circular gaps ($\sim$1.7 mm in diameter) in confluent monolayers of mouse embryonic epicardial cells (MEC1) under two conditions, without and with TGF-$\beta$ treatment. We show that both tissue fluidity and the elastic properties associated with fiber reinforcement are critical for reproducing the closure kinematics. Specifically, we predict that the TGF-$\beta$ treated condition has lower fluidity, associated with a lower fluidic deformation rate and a higher elastic deformation rate than the untreated condition, in agreement with the experimental observations.
\end{abstract}


\end{frontmatter}
\section{Introduction}
Closed-border gap closure in confluent epithelial tissues involves inward tissue migration that fills an enclosed open space. Understanding the mechanics of gap closure is critical for manipulating the process. Both {\it in vivo} and {\it in vitro} experiments have demonstrated that actin-cable contractility and cell crawling at the wound border of animal epithelial tissues are key mechanical factors for rapid closure; see \citep{martin1992actin,jacinto2002dynamic,anon2012cell,brugues2014forces}, and the review by \citet{begnaud2016mechanics}. More recently, increasing attention has been paid to fluidization due to cell–cell intercalation near the wound border in embryonic animal tissues \citep{tetley2019tissue,jain2020regionalized}.  From a geometric perspective, cell–cell intercalation allows cells near the border to remain more or less packed (rather than elongated) as the gap constricts. 

In the context of polarized or pulsatile actomyosin contractility–driven morphogenesis, such cell–cell intercalation is understood as “active” or internal-force–driven \citep{razzell2014recapitulation,collinet2015local}. While this process requires local active forces, when tissue-scale forces due to proliferation \citep{mao2013differential}, anisotropic compression \citep{collinet2015local}, or tension \citep{butler2009cell,pare2014positional} are at play, it can be understood as a process that further relaxes the tissue elastic energy \citep{farhadifar2007influence,rauzi2008nature}. For static tissues, reducing actomyosin contractility promotes cell–cell intercalation \citep{curran2017myosin}, which can be understood as a transition from a solid to a fluid state \citep{bi2014energy,bi2015density}.

In this manuscript, along with \citet{Jiangplaceholder2027placeholder}, we report a novel growth-driven closure dynamics for large circular gaps ($\sim$1.7 mm in diameter) in confluent monolayers made of mouse embryonic epicardial cells (MEC1). MEC1 cell monolayers present different cellular and supracellular organizations of actin fibers compared with previously studied epithelial monolayers due to their “partial EMT” phenotype. In short, while staying connected with their neighbors, each cell presents medial actin fibers resembling the organization of mesenchymal cells, whereas along the cell–cell junctions the directional alignment between the interface direction and the fibers is weakened or absent. Strikingly, even though the entire closing process took about one day due to the large initial gap size, the closing dynamics remained more or less circular, suggesting a highly mechanically regulated process. See counterexamples \citep{ben2014re,Jiangplaceholder2027placeholder} for complications in shape development during Madin–Darby canine kidney (MDCK) cell gap closure. Associated with the quasi-circular closure, we observed large-scale tissue growth, extension–contraction, and actin fiber alignment along the local radial direction beyond the border region during the process. At the late stage, we observed an area constriction in the bordering regions. Among the control and TGF-$\beta$1–treated conditions along the partial EMT spectrum (with more EMT in the treated case), the treated case closed slightly faster, with a smaller drop in migration speed away from the wound front, and presented more exaggerated tissue-level actin fiber alignment along the radial direction. At the cell level, we did not observe more cell division but did observe less cell–cell intercalation in the rear region, associated with radial cell elongation. 

To elucidate the closing process and the mechanical differences between the two conditions, we develop a continuum model. We consider the MEC1 monolayer as a 3D incompressible hyperelastic material with actin fiber reinforcement limited to the in-plane directions. We model the growth and fluidization through growth-elasticity theory \citep{Rodriguez1994}, where the deformation is decomposed into elastic and inelastic parts. In the Eulerian frame, the evolution of the elastic deformation can be described by a tensorial growth rate and a deviatoric relaxation rate. We model the fluidization due to cell–cell intercalation through the deviatoric relaxation rate, which extends or contracts according to the 3D stress state in a mass-conserving fashion. Similar approaches have been considered in growth-elasticity–based wound-closure modeling \citep{taber2009towards,bowden2016morphoelastic}, where the local stress state guides the rearrangements. By contrast, our derivation identifies tissue rearrangement as an elastic-energy relaxation mechanism within the monolayer \citep{farhadifar2007influence,mao2013differential}. This formulation also draws a direct connection to active-fluid models used for gap closure \citep{ravasio2015gap} and convergent extension that involving extension--contraction deformations \citep{dicko2017geometry,streichan2018global,ioratim2023mechanochemical}. The model further incorporates actin-fiber reinforcement in the bulk, motivated by the actin organization observed in MEC1 monolayers. In our previous work \citep{olaranont2025chemomechanical}, we also developed a Maxwell-type model by associating the rearrangement with the right Cauchy-Green elastic deformation for neo-Hookean materials. The model developed in this manuscript directly connects rearrangement with the Cauchy stress, which works with any hyperelastic material models. Specifically, we incorporate planar fiber reinforcement into the 3D neo-Hookean thin-plate material, appropriate for cell monolayers. Finally, we demonstrate how the model can be integrated with experimental kinematic data, where it decomposes the radial tissue flow into growth-related, elastic, and fluidic subflows.


The manuscript is organized as follows. In Section~\ref{Model}, we develop a continuum model for a growing, fluidizing, fiber-reinforced tissue monolayer, derive its reduced thin-plate form, and specialize the model to the radially symmetric geometry used for circular-gap closure. In Section~\ref{Result}, we first describe the MEC1 closure experiments, including the partial-EMT actin organization, inward tissue flow, extension--contraction, and late-stage areal constriction. We then fit the model to the experimental kinematics under the control and treated conditions and use the fitted simulations to illustrate the contributions of fluidic rearrangements and fiber reinforcement to regulating wound-closure dynamics and mechanics. In Section~\ref{Discussion}, we summarize the results in the context of the current understanding of tissue flow dynamics. Additional experimental data and model features that help elucidate the experiments, as well as comparisons with previous wound-closure studies, are presented in \citet{Jiangplaceholder2027placeholder}.

\section{Continuum modeling of a growing plate with fluidization and fiber-reinforcement}\label{Model}

\subsection{Three-dimensional model of growing tissue monolayer}
\label{Subsec.Model_elastic_3D}
Consider the tissue monolayer occupying the region
$\mathcal{B} =\{(x,y,z)\,:\,\mathbf{x}=(x,y)\in\Omega_t,\ 0<z<h(x,y,t)\}$  at time $t$, where $h(x,y,t)$ is the thickness of the monolayer. 
The total strain (elastic) energy of tissues stored within $\mathcal{B}$ is given by
\begin{equation}
E_w(t)=\int_{\Omega_t}\int_0^{h(x,y,t)} W(\mathbf{F}_e) J_e^{-1} \mathrm{d}z\mathrm{d}\mathbf{x},
\end{equation}
where $\mathbf{F}_e$ is the elastic deformation tensor, $W(\mathbf{F}_e)$ represents the strain energy density per stress-free volume, $J_e = \det(\mathbf{F}_e)$ is the volumetric variation (ratio) due to elastic deformation, and hence $W(\mathbf{F}_e)J_e^{-1}$ represents the strain energy density per deformed volume in the Eulerian frame. The rate of change of $E_w(t)$ is 
\begin{align}\label{eq:dEw_dt}
\frac{d E_w}{dt} &= \frac{d}{dt}\int_{\Omega_t}\int_0^{h} J_e^{-1}W(\mathbf{F}_e)\mathrm{d}z\mathrm{d}\mathbf{x} = \int_{\Omega_t}\int_0^{h} \Big(\frac{d}{dt}(J_e^{-1}W) + J_e^{-1}W \nabla \cdot \mathbf{v} \Big) \mathrm{d}z\mathrm{d}\mathbf{x}, \nonumber  \\  
&= \int_{\Omega_t}\int_0^{h} \Big(J_e^{-1}\frac{\partial W}{\partial \mathbf{F}_e}:\frac{d\mathbf{F}_e}{dt} -J_e^{-2}\frac{dJ_e}{dt} W+ J_e^{-1}W \nabla \cdot \mathbf{v} \Big) \mathrm{d}z\mathrm{d}\mathbf{x},  
\end{align}
where we have used the Reynolds transport theorem in the first equality, $\mathbf{v}$ is the velocity field, and $\nabla \cdot \mathbf{v}$ represents rate of volume expansion/contraction. By extending the growth-elasticity theory \citep{Rodriguez1994,Goriely2009-morpho-elastic} to Eulerian frame \citep{yan2021stress,Wei2023elasticmodel,olaranont2025chemomechanical}, the dynamics of the elastic deformation tensor is given by
\begin{align}\label{eq:Fe_evol}
\frac{d \mathbf{F}_e}{dt}  = (\nabla\mathbf{v}-\boldsymbol{\Gamma})\mathbf{F}_e,\qquad
\frac{d J_e}{dt}  = \left(\nabla\cdot \mathbf{v}- \mathrm{tr}(\boldsymbol{\Gamma})\right)J_e,
\end{align}
where $\boldsymbol{\Gamma}$ is the rate of inelastic deformation due to tissue growth and rearrangement. Note that the dynamics \eqref{eq:Fe_evol} reduces to the classical relation $d\mathbf{F}/dt =\nabla\mathbf{v} \mathbf{F}$ without inelastic deformation for the (geometric) deformation gradient. Moreover, we decompose the rate of inelastic deformation into the pure growth part and the rearrangement part as
\begin{equation}
\boldsymbol{\Gamma} = \boldsymbol{\Gamma}_g + \boldsymbol{\Gamma}_D,
\end{equation}
where $\boldsymbol{\Gamma}_g$ is the pure growth rate tensor due to cell proliferation and apoptosis, whereas $\boldsymbol{\Gamma}_D$ accounts for the isochoric rearrangement due to local remodeling of tissues, assumed to be traceless, i.e., $\mathrm{tr}(\boldsymbol{\Gamma}_D)=0$. 

Assume that the Cauchy stress only depends only on the elastic deformation, i.e., $\boldsymbol{\sigma} = \boldsymbol{\sigma}(\mathbf{F}_e)$, we have the following constitutive relation:	
\begin{align} \label{stress}
\boldsymbol\sigma = {J_e}^{-1}\frac{\partial W}{\partial \mathbf{F}_e} \mathbf{F}_e^{\top}.
\end{align}
Moreover, we assume that the fluidic behavior of tissue mechanics due to rearrangement is described via the dependence of the evolution of $\mathbf{F}_e$ on the isochoric rearrangement rate $\boldsymbol{\Gamma}_D$ \eqref{eq:Fe_evol}. In particular, we assume that the thermodynamically consistent rearrangement is an energy-dissipative process:
\begin{equation}
\boldsymbol{\sigma}:\boldsymbol{\Gamma}_D\geq 0.
\end{equation}
Inserting \eqref{stress} and \eqref{eq:Fe_evol} into \eqref{eq:dEw_dt}, we obtain the following:
\begin{align}\label{eq:Ewdynamic}
\frac{d E_w}{dt} = \int_{\Omega_t}\int_0^{h} \Big(\boldsymbol{\sigma}: \nabla \mathbf{v} - \boldsymbol{\sigma}:\boldsymbol{\Gamma} +  J_e^{-1}W \mathrm{tr}(\boldsymbol{\Gamma}_g)\Big) \mathrm{d}z\mathrm{d}\mathbf{x}, 
\end{align}
where the three terms on the right-hand side represent, respectively, the stress power due to total (geometric) deformation, the stress power due to inelastic deformation, and the addition (loss) of elastic energy due to volumetric growth (loss). 

Using the relation
$\boldsymbol{\sigma}:\nabla \mathbf{v} = \nabla\cdot(\boldsymbol{\sigma}\cdot\mathbf{v}) - (\nabla\cdot \boldsymbol{\sigma})\cdot\mathbf{v}$ and the divergence theorem, we can rewrite the stress power due to deformation as
\begin{align}\label{eq:power_decomp}
\int_{\Omega_t}\int_0^{h} \boldsymbol{\sigma}:\nabla \mathbf{v} \mathrm{d}z\mathrm{d}\mathbf{x} 
= &
-\int_{\Omega_t}\int_0^{h(x,y,t)} (\nabla\cdot \boldsymbol{\sigma})\cdot\mathbf{v}\,\mathrm{d}z\,\mathrm{d}\mathbf{x} \\
&+\int_{\partial\Omega_t} \int_0^{h(x(s),y(s),t)}(\boldsymbol{\sigma}\cdot\mathbf{n}_{||})\cdot\mathbf{v}\,\mathrm{d}z\mathrm{d}s \\
&+\int_{\Omega_t}(\boldsymbol{\sigma}\cdot\tilde{\mathbf{n}}_{\mathrm{top}})\cdot\mathbf{v}(x,y,h(x,y),t)\,\mathrm{d}\mathbf{x} \\
&-\int_{\Omega_t}(\boldsymbol{\sigma}\cdot\mathbf{e}_{z})\cdot\mathbf{v}(x,y,0,t)\,\mathrm{d}\mathbf{x} 
\end{align}
where in the second line $\mathbf{n}_{||}$ denotes the outward unit normal to the lateral surface ${\partial\mathcal{B}}_{||}=\{(x,y,z): \mathbf{x}=(x,y)\in\partial\Omega_t, 0\leq z \leq h(x(s),y(s),t)\}$ with $s$ being the arclength of $\partial\Omega_t$, in the third line $\tilde{\mathbf{n}}_{\mathrm{top}}=(-h_x,-h_y,1)$ denotes the outward normal to the top surface rescaled by the areal factor $\sqrt{1 + h_x^2 + h_y^2}$, and in the fourth line $\mathbf{e}_{z}=(0,0,1)$.

We further assume that the tissue monolayer is in mechanical equilibrium:
\begin{equation}\label{eq:mech_equil}
\nabla \cdot \boldsymbol{\sigma} = \mathbf 0,~~\rm{in}~~\mathcal{B},
\end{equation}
with mixed boundary conditions:
\begin{align}
\label{eq:BC_top}
\boldsymbol{\sigma}\cdot \tilde{\mathbf{n}}_{\mathrm{top}} &= \mathbf0, \quad && \text{at traction-free top surface ${\partial\mathcal{B}}_{\mathrm{top}}$},\\
\label{eq:BC_bottom}
\mathbf{v}(x,y,0,t)\cdot\mathbf{e}_{z} &= 0, \quad && \text{at fixed bottom surface ${\partial\mathcal{B}}_{\mathrm{bottom}}$}, \\
\label{eq:BC_traction} 
\boldsymbol{\sigma}\cdot \mathbf{n}_{||} &= \mathbf{F}_{\mathrm{ext}}, \quad && \text{at moving lateral boundary $\partial\mathcal{B}_{||}\backslash\partial\mathcal{B}_{||C}$},\\
\label{eq:fix_bc}
\mathbf{v} &= \mathbf0, \quad && \text{at fixed lateral boundary $\partial\mathcal{B}_{||C}$},
\end{align}
where $\mathbf{F}_{\mathrm{ext}}$ represents external traction, which turns out to be negligible compared to internal stresses during the circular wound closure in MEC1 monolayers, i.e., $\mathbf{F}_{\mathrm{ext}}= \mathbf{0}$.

\subsection{Scale separation of mechanical stresses in the thin monolayer}
\label{Subsec.leading_order_approximation}
Next, we show the scale separation between the in-plane and out-of-plane stresses in the thin monolayer by taking advantage of its thin-plate geometry. Assume that the thickness of tissue layer is very small compared to the in-plane characteristic length scale $L$, i.e., $h/L = O(\varepsilon)$, and the top surface has small slopes, i.e., $|\nabla_\textbf{x} h|=O(\varepsilon)$ ($\mathbf{x}=(x,y)$), with a small parameter $0<\varepsilon\ll 1$. Denoting the Cauchy stress $\boldsymbol{\sigma}=\{\sigma_{ij}\}_{i,j=1,2,3}$, we further assume the planar stresses satisfy $\sigma_{ij}=O(1)$ for $i,j=1,2$ and the in-plane variations of all stress is in agreement with the length scale $L$, i.e., $\partial_x\sigma_{ij}, \partial_y\sigma_{ij}=O(\sigma_{ij}/L)$ for $i,j=1,2,3$. The traction-free boundary condition at the top surface \eqref{eq:BC_top} yields the relation
\begin{equation}\label{eq:traction-identity}
\sigma_{i3}=h_x\,\sigma_{i1}+h_y\,\sigma_{i2},
\qquad i=1,2,3,
\qquad \text{on } z=h(x,y).
\end{equation}
Given that $h_x,h_y=O(\varepsilon)$, we directly obtain $\sigma_{13}(x,y,h), \sigma_{23}(x,y,h)=O(\varepsilon)$. By inserting $\sigma_{13}(x,y,h), \sigma_{23}(x,y,h)$ into the right-hand side of \eqref{eq:traction-identity}, we further obtain $\sigma_{33}(x,y,h)=O(\varepsilon^2)$.

On the other hand, the mechanical equilibrium condition \eqref{eq:mech_equil} yields 
\begin{equation}\label{eq:divsigma-components}
\partial_z \sigma_{i3}=-\partial_x \sigma_{i1}-\partial_y \sigma_{i2},
\qquad i=1,2,3,
\qquad \text{in $\mathcal{B}$}.
\end{equation}
Integrating along $z$ direction, we obtain the following estimate
\begin{align*}
|\sigma_{i3}(x,y,z)|&=|\sigma_{i3}(x,y,h)-\int_{z}^{h(x,y)} \partial_z \sigma_{i3}(x,y,\xi)\,\mathrm{d}\xi | \\
&\leq |\sigma_{i3}(x,y,h)| + h(x,y) \cdot \mathrm{max}_{\xi}|\partial_x \sigma_{i1}+\partial_y \sigma_{i2}| \\
&\leq O(|\sigma_{i3}(x,y,h)|) + \varepsilon \cdot (O(|\sigma_{i1}|)+O(|\sigma_{i2})|)
\end{align*}
where we have used $\partial_x\sigma_{ij},\partial_y\sigma_{ij}=O(\sigma_{ij}/L)$ and $h/L=O(\varepsilon)$ in the last line. Therefore, we obtain the estimate $\sigma_{13},\sigma_{23}=O(\varepsilon)$ and $\sigma_{33}=O(\varepsilon^2)$ in the bulk. In summary, the stress tensor is given by
\begin{equation}\label{eq:sigma_mtx}
	\boldsymbol{\sigma} =
	\begin{bmatrix}
		\boldsymbol{\sigma}_{||} &  O(\varepsilon) \\
		O(\varepsilon)           &  O(\varepsilon^2)
	\end{bmatrix}, \qquad
	\boldsymbol{\sigma}_{||} =
	\begin{bmatrix}
	\sigma_{11} & \sigma_{12} \\
	\sigma_{12} & \sigma_{22} \\
	\end{bmatrix}.
\end{equation}

\subsection{Thermodynamically-consistent modeling of tissue rearrangement in thin monolayer}
\label{Subsec:modeling_rearrangement}
Now, combining $\sigma_{i3}=O(\varepsilon)$ (for $i=1,2$) with $v_3(x,y,0,t)=0$ derived from \eqref{eq:BC_bottom}, the stress power due to deformation \eqref{eq:power_decomp} vanishes at leading order:
\begin{equation}\label{eq:power_reduce}
\int_{\Omega_t}\int_0^{h} \boldsymbol{\sigma}:\nabla \mathbf{v} \mathrm{d}z\mathrm{d}\mathbf{x} 
=-\int_{\Omega_t}(\boldsymbol{\sigma}\cdot\mathbf{e}_{z})\cdot\mathbf{v}(x,y,0,t)\,\mathrm{d}\mathbf{x} = O(\varepsilon).
\end{equation}
where we have used the mechanical equilibrium \eqref{eq:mech_equil} and boundary conditions \eqref{eq:BC_top}--\eqref{eq:fix_bc}. 
Moreover, the stress power due to inelastic deformation reduces to 
\begin{align}\label{eq:power_growth}
	\int_{\Omega_t}\int_0^{h} \boldsymbol{\sigma}:\boldsymbol{\Gamma} \,\mathrm{d}z\mathrm{d}\mathbf{x} &=  
	h \int_{\Omega_t} \boldsymbol{\sigma}:\left(\boldsymbol{\Gamma}_{g}
	+ \boldsymbol{\Gamma}_{D}\right)\Big\rvert_{z=h/2} \,\mathrm{d}\mathbf{x} + O(\varepsilon^2), \nonumber \\
	&= h \int_{\Omega_t} \left(\gamma\mathrm{tr}(\boldsymbol{\sigma}_{||})
	+ \boldsymbol{\sigma}_{D}:\boldsymbol{\Gamma}_{D}\right)\Big\rvert_{z=h/2} \,\mathrm{d}\mathbf{x} + O(\varepsilon^2), 
\end{align}
where we assume isotropic planar areal growth $\boldsymbol{\Gamma}_{g}=\mathrm{diag}(\gamma,\gamma,0)$ with a growth rate $\gamma$ due to cell division within the monolayer, and the traceless deviatoric stress is:
\begin{equation}\label{sigma_D}
\boldsymbol{\sigma}_{D}= \boldsymbol{\sigma}-\frac{1}{3}\Tr\left(\boldsymbol{\sigma}\right)
= \mathrm{diag}\left(\boldsymbol{\sigma}_{||}-\frac{1}{3}\Tr(\boldsymbol{\sigma}_{||})\mathbf{I}_2, -\frac{1}{3}\Tr(\boldsymbol{\sigma}_{||})\right)+O(\varepsilon).
\end{equation}
Hereinafter, we will restrict our discussion to the leading-order approximation, neglecting higher-order terms.

Following \citet{olaranont2025chemomechanical}, we assume that the local isochoric rearrangement to be an energy-dissipative action such that $\boldsymbol{\sigma}:\boldsymbol{\Gamma}_{D}=\boldsymbol{\sigma}_{D}:\boldsymbol{\Gamma}_{D} \geq 0$, which suggests that $\boldsymbol{\Gamma}_D$ relates to the deviatoric part of stress. However, unlike the spherical geometry for tumor growth in \citet{olaranont2025chemomechanical}, we consider a thin-plate geometry for wound closure, which suggests that the cell--cell intercalation is predominant within the plane while the $z$-dimension is not involved. Therefore, denoting $\beta\geq 0$ and $\zeta\geq 0$ the intercalation specific and non-intercalation specific fluidity (inverse of viscosity), respectively, we assume the following form of $\boldsymbol{\Gamma}_D$:
\begin{equation}\label{eq:GammaD}
	\boldsymbol{\Gamma}_{D} = \beta\boldsymbol{\sigma}_{D_{||}} +\zeta\boldsymbol{\sigma}_{D},
\end{equation}
where $\boldsymbol{\sigma}_{D_{||}}$ represents the in-plane deviatoric stress due to the planar cell-intercalation:
\begin{equation}
\boldsymbol{\sigma}_{D_{||}} =
\begin{bmatrix}
\boldsymbol{\sigma}_{||}-\frac{1}{2}\Tr(\boldsymbol{\sigma}_{||})\mathbf{I}_2 &  0 \\
0            &  0
\end{bmatrix}.
\end{equation}
Moreover, we further introduce the out-of-plane deviatoric stress $\boldsymbol{\sigma}_{D_{\perp}}$ such that $\boldsymbol{\sigma}_{D}=\frac{2}{3}\boldsymbol{\sigma}_{D_{||}}+\frac{1}{3}\boldsymbol{\sigma}_{D_{\perp}}$:
\begin{equation}
\boldsymbol{\sigma}_{D_{\perp}} =
\begin{bmatrix}
\boldsymbol{\sigma}_{||} &  0 \\
0   &  -\Tr(\boldsymbol{\sigma}_{||})
\end{bmatrix}.
\end{equation}
Then the tissue arrangement $\boldsymbol{\Gamma}_D$ can be decomposed as the planar and out-of-plane arrangement parts:
\begin{equation}\label{eq:GammaD2}
	\boldsymbol{\Gamma}_{D} = (\beta+\frac{2}{3}\zeta)\boldsymbol{\sigma}_{D_{||}}+\frac{1}{3}\zeta\boldsymbol{\sigma}_{D_{\perp}}.
\end{equation}
One can easily check that above choice of $\boldsymbol{\Gamma}_{D}$ guarantees energy dissipation
\begin{equation}
\boldsymbol{\sigma}_{D}:\boldsymbol{\Gamma}_{D} = (\beta+\zeta)\boldsymbol{\sigma}_{D_{||}}:\boldsymbol{\sigma}_{D_{||}}+\frac{\zeta}{6}\left(\Tr(\boldsymbol{\sigma}_{||})\right)^2 \geq 0.
\end{equation}

\subsection{Incompressible Neo-Hookean elasticity with fiber-reinforcement}
\label{Subsec.constitutive_law}
The above discussion on the mechanics of tissue monolayer is valid for general constitutive of elasticity. Now we assumed that the monolayer is a thin sheet of incompressible neo-Hookean material with planar fiber-reinforcement: 
\begin{equation}\label{eq:W}
W(\mathbf{F}_e) = \frac{\mu}{2}({\mathbf{F}_e^\top\mathbf{F}_e}:\mathbf{I}-3)+\frac{1}{4}\eta(\mathbf{F}_e^\top\mathbf{F}_e:\mathbf{H}-1)^2 -p(\det{\mathbf{F}_e}-1),
\end{equation}
where $\eta$ is the elastic modulus due to the fiber reinforcement relative to the ground material shear modulus, $\mathbf{H}=\mathrm{diag}(k_H,1-k_H,0)$ is the fiber structure tensor with $k_H$ describing the local dispersion level of fiber orientation, and $p$ is the pressure to ensure incompressibility in 3D. 
Then the Cauchy stress tensor is given by \eqref{stress}
\begin{equation}\label{eq:sigma3d_exp}
\boldsymbol\sigma =  \mu \mathbf{F}_e\mathbf{F}_e^{\top}+\eta (I_k-1)\mathbf{F}_e\mathbf{H}\mathbf{F}_e^\top-p\mathbf{I} 
\end{equation}
where $I_k=\mathbf{F}_e^\top\mathbf{F}_e:\mathbf{H}$. 

Since $|\nabla_{\mathbf{x}}h|=O(\varepsilon)$, the deformation gradient $\mathbf{F}_e=\mathrm{diag}(\mathbf{F}_{e_{||}},{F_e}_{33})$ is block-diagonal at leading order, where $\mathbf{F}_{e_{||}}$ is the planar elastic deformation tensor. Inserting the constitutive relation \eqref{eq:sigma3d_exp} into \eqref{eq:sigma_mtx} yields $\sigma_{33}=\mu ({F_e}_{33})^2-p=0$. Combining with the incompressibility condition $\det(\mathbf{F}_e)=1$, we can solve pressure in terms of planar strains:
\begin{equation}\label{pressure}
p=\mu \left({F_e}_{33}\right)^2=\mu \left(1/\det(\mathbf{F}_{e_{||}})\right)^2 .
\end{equation}
Inserting the above result back into \eqref{eq:W} and \eqref{eq:sigma3d_exp}, the mechanics of tissue monolayer is solely determined by planar deformation. 

\subsection{Reduced two-dimensional mechanical model of tissue monolayer}
With the above derivation and discussion, the 3D mechanical model of tissue monolayer can be reduced to a quasi-2D system: 
\begin{align}\label{2Dsystem}
\begin{dcases}
\text{Mechanical Equilibrium: } \\
\nabla_{\mathbf{x}} \cdot \boldsymbol{\sigma}_{||} = \mathbf 0,  \\
\text{Elastic Deformation Dynamics: } \\
\frac{d \mathbf{F}_{e_{||}}}{dt} = \left(\nabla_{\mathbf{x}}\mathbf{v}_{||}-\left(\gamma +\frac{\zeta}{6}\mathrm{tr}(\boldsymbol{\sigma}_{||})\right)\mathbf{I}_2\right)\mathbf{F}_{e_{||}}-(\beta+\zeta)\boldsymbol{\sigma}_{{||}_D}\mathbf{F}_{e_{||}},
\end{dcases}
\end{align}
subject to the initial conditions 
\begin{equation}\label{2DIC}
\mathbf{F}_{e_{||}}(\mathbf{x},0) = \mathbf I_2,
\end{equation}
and mixed boundary conditions
\begin{align}\label{2DBC}
\begin{dcases}
\boldsymbol{\sigma}_{||}\cdot \mathbf{n}_{||} = \mathbf{0}, \quad & \text{at a moving part of the boundary $\Sigma_t=\partial\Omega_t\backslash\Sigma_C$},\\
\mathbf{v}_{||} = \mathbf0, \quad & \text{at a fixed part of the boundary $\Sigma_C$},
\end{dcases}
\end{align}

\subsection{Deviatoric Stress Relaxation and Maxwell-Type Limit}
\label{subsec:Maxwell}
Next we demonstrate the stress-relaxation behavior in the linearization of our nonlinear model. In the theory of small deformations, we assume that $\mathbf{F}_{||} = \mathbf{I}_2 + \nabla_{\mathbf{x}} \mathbf{u}$ and $\mathbf{F}_{e_{||}} = \mathbf{I}_2 + \mathbf{E}$, where the planar displacement $\mathbf{u} = \mathbf{x}-\mathbf{X}$ and the elastic increment tensor $\mathbf{E}$ are small such that nonlinear terms can be discarded.

The linearized planar Cauchy stress is given by:
\begin{equation}
	\boldsymbol{\sigma}_{||} = \mu \left(\mathbf{E}+\mathbf{E}^\top\right) +\eta \left(\left(\mathbf{E}+\mathbf{E}^\top\right):\mathbf{H}_{||}\right)\mathbf{H}_{||}-p\,\mathbf{I}_2,
\end{equation}
where $\mathbf{H}_{||}=\textbf{diag}(1-k_H, k_H)$ represents the planar fiber orientation tensor. Under small elastic deformations, the dynamics of elastic increment tensor is 
\begin{equation}
\dot{\mathbf{E}}=\nabla_{\mathbf{x}} \mathbf{v}_{||}- \boldsymbol{\Gamma}_{||},
\end{equation}
which gives the evolution of linearized stress
\begin{equation}
\begin{aligned}
	\dot{\boldsymbol{\sigma}}_{||}
	= 2\mu \mathbf{D}_e
	+ 2\eta(\mathbf{D}_e:\mathbf{H}_{||})\mathbf{H}_{||} 
	- \dot{p}\,\mathbf{I}_2,
\end{aligned}
\end{equation}
where $\dot{p}= -2\mu\left(\nabla_{\mathbf{x}}\cdot\mathbf{v}_{||}
	-\left(2\gamma+\frac{\zeta}{3}\Tr(\boldsymbol{\sigma}_{||})\right)\right)$ from \eqref{pressure} and we define
\begin{equation}\label{eq:strain_rate}
\mathbf{D}_e = \mathbf{D}-\mathbf{D}_\Gamma, \quad \mathbf{D}=\tfrac12(\nabla_{\mathbf{x}} \mathbf{v}_{||}+\nabla_{\mathbf{x}} \mathbf{v}_{||}^{\mathsf T}), 
\quad
\mathbf{D}_{\Gamma}=\tfrac12(\boldsymbol{\Gamma}_{||}+\boldsymbol{\Gamma}_{||}^{\mathsf T}).
\end{equation}
Taking the isotropic part and deviatoric part, we obtain
\begin{align}
\label{eq:anisotropic_maxwell}
&\dot{\boldsymbol{\sigma}}_{||_p}
= 3\mu \left(\nabla_{\mathbf{x}} \cdot \mathbf{v}_{||}-(2\gamma+\tfrac{\zeta}{3}\Tr(\boldsymbol{\sigma}_{||}))\right)\mathbf{I}+\eta \left(\mathbf{D}_e:\mathbf{H}_{||}\right)\mathbf{I},\\
&\dot{\boldsymbol{\sigma}}_{||_D} + 2\mu (\beta+\zeta)\boldsymbol{\sigma}_{||_D} 
=  2\mu \mathbf{D}_D +2\eta \left(\mathbf{D}_e:\mathbf{H}_{||}\right)\mathbf{H}_{||_D},
\end{align}
where we define
\begin{equation}
\mathbf{D}_D = \tfrac{1}{2}\left(\nabla_{\mathbf{x}} \mathbf{v}_{||}+\nabla_{\mathbf{x}} \mathbf{v}_{||}^{\mathsf T} - \left(\nabla_{\mathbf{x}} \cdot \mathbf{v}_{||}\right)\mathbf{I}_2 \right), \quad 
\mathbf{H}_{||_D} = \mathbf{H}_{||}-\tfrac{1}{2}\mathbf{I}_2
\end{equation}
When the fiber alignment is isotropic ($k_H=1/2$ and $\mathbf{H}_{||_D}=\mathbf{0}$) or the fiber reinforcement is absent ($\eta=0$), \eqref{eq:anisotropic_maxwell} simplifies to the classical linear Maxwell model:
\begin{equation}\label{eq:maxwell_classical}
	\dot{\boldsymbol{\sigma}}_{||_D}
	+ \frac{1}{\tau}\boldsymbol{\sigma}_{||_D}
	= 2\mu\,\mathbf{D}_D,
	\qquad
	\tau = \frac{1}{2\mu (\beta+\zeta)}.
\end{equation}

\subsection{Radial symmetric systems}
Consider the system in cylindrical coordinates with radial symmetry in $xy$-plane. We have
\begin{align}
&\mathbf{F}_e = \textbf{diag}({f_e}_r,{f_e}_\theta,{f_e}_z), \quad \mathbf{H}=\textbf{diag}(1-k_H, k_H, 0),\\
&\mathbf{v} = (v,0,v_z)^{\text{T}}, \quad
\nabla \mathbf{v} = \textbf{diag}(\frac{\partial v}{\partial r}, \frac{v}{r},\frac{\partial v_z}{\partial z}), \quad
\boldsymbol\sigma = \textbf{diag}(\sigma_{rr},\sigma_{\theta\theta},0), \\
&\boldsymbol{\Gamma}_g = \textbf{diag}(\gamma, \gamma, 0), \quad
\boldsymbol{\Gamma}_D = \textbf{diag}(\gamma_{Dr}, \gamma_{D\theta}, -(\gamma_{Dr}+\gamma_{D\theta})).
\label{arealrearrangment}
\end{align}
The 2D system \eqref{2Dsystem}-\eqref{2DBC} reduces to: 
\begin{equation}\label{eq:radial_system}
\begin{dcases}
&\frac{\partial {{f_e}_r}}{\partial t}+v\frac{\partial {{f_e}_r}}{\partial r}=\left(\frac{\partial v}{\partial r}-\gamma-\gamma_{Dr}\right){{f_e}_r}\\   
&\frac{\partial {{f_e}_\theta}}{\partial t}+v\frac{\partial {{f_e}_\theta}}{\partial r}=\left(\frac{v}{r}-\gamma-\gamma_{D\theta}\right){{f_e}_\theta} \\
&\frac{\partial \sigma_{rr}}{\partial r}+\frac{1}{r}(\sigma_{rr}-\sigma_{\theta\theta})=0,\\
&\sigma_{rr}= \big(\mu+\eta(1-k_H)(I_k-1)\big){f_e}_r^2-\mu {f_e}_r^{-2}{f_e}_\theta^{-2}, \\ 
&\sigma_{\theta\theta}= \big(\mu+\eta k_H(I_k-1)\big){f_e}_\theta^2-\mu {f_e}_r^{-2}{f_e}_\theta^{-2}, \\
&\gamma_{Dr} = \frac{\beta}{2}(\sigma_{rr}-\sigma_{\theta\theta})+\frac{\zeta}{3}(2\sigma_{rr}-\sigma_{\theta\theta}),\\
&\gamma_{D\theta} = \frac{\beta}{2}(\sigma_{\theta\theta}-\sigma_{rr})+\frac{\zeta}{3}(2\sigma_{\theta\theta}-\sigma_{rr}),\\
\end{dcases}	
\end{equation}
where $I_k = (1-k_H){f_e}_r^2+k_H{f_e}_\theta^2$ and subject to the initial and boundary conditions
\begin{equation} \label{eq:bc}
\begin{dcases}
{f_e}_r(r,0)={f_e}_\theta(r,0)=1, \quad R(0)=R_0, \quad \text{at $t=0$}\\
\sigma_{rr}(R,t)=0,\quad \frac{dR}{dt} = v(R,t), \quad \text{at $r=R(t)$}, \\
v(R_{out},t)=0,  \quad \text{at $r=R_{out}$}\\
\end{dcases}
\end{equation}
Notice that the 3D dynamics is enslaved to the solution 
of the 2D system, where the out-of-plane deformation is given by ${f_e}_z = 1/({f_e}_r{f_e}_\theta)$ and the out-of-plane velocity satisfies
\begin{equation}
\frac{\partial v_z}{\partial z}=2\gamma-\frac{\partial v}{\partial r}-\frac{v}{r}, \qquad
v_z(r,z,t)=\int_{0}^{z}\left(2\gamma-\frac{\partial v}{\partial r}-\frac{v}{r}\right)(r,\xi,t)\,\mathrm{d}\xi .
\end{equation}

To summarize, our model describes how the global velocity and stress is patterned by a global planar growth rate $\gamma$ and the fluidic tendency considering both planar fluidity $\beta$ due to cell intercalation and thickness fluidity $\zeta$. See Sec.~\ref{append:numerical} for the numerical methods.

\section{Results}\label{Result}
\subsection{Gap closure in MEC1 cell monolayers}
\subsubsection{Confluent tissue with a partial EMT-phenotype actin network}
We placed cylindrical stencils with a diameter of $1.7$ mm to manufacture large circular gaps in embryonic epicardial cell (MEC1) monolayers. See \citet{Jiangplaceholder2027placeholder} for experimental protocols. The MEC1 monolayers were confluent, but the individual cells had less definition of apical--basal polarity compared to epithelial cells, due to their partial epithelial-to-mesenchymal transition (pEMT) phenotype \citep{Jiangplaceholder2027placeholder}. While these individual cells have lost apical--basal polarity (to a varying extent) compared to epithelial cells, they still remained connected with their neighbors and formed tissue-level actin networks (Fig.~\ref{Fig1S}a, control), except that an insignificant fraction of front boundary cells detached from the global monolayer.

Unlike in typical epithelial tissues where actin fibers are organized along the cell--cell junctions, actin fibers in MEC1 monolayers presented different intracellular and intercellular organizations: within the cells, the fibers resembled a more mesenchymal organization with medial actin fibers, whereas along the cell--cell junctions the directional alignment between the junction direction and the fibers was weakened and more dispersed (Fig.~\ref{Fig1S}a, control, Box A). As closure continued, patches of tissue-wide cell and actin radial alignment emerged. These radial alignments became more obvious when the tissue was induced toward a more mesenchymal phenotype along the pEMT spectrum by TGF-$\beta$ treatment (Fig.~\ref{Fig1S}a, treated, Box B) \citep{Jiangplaceholder2027placeholder}.

\begin{figure}[htbp]
\centering
\includegraphics[width=\textwidth]{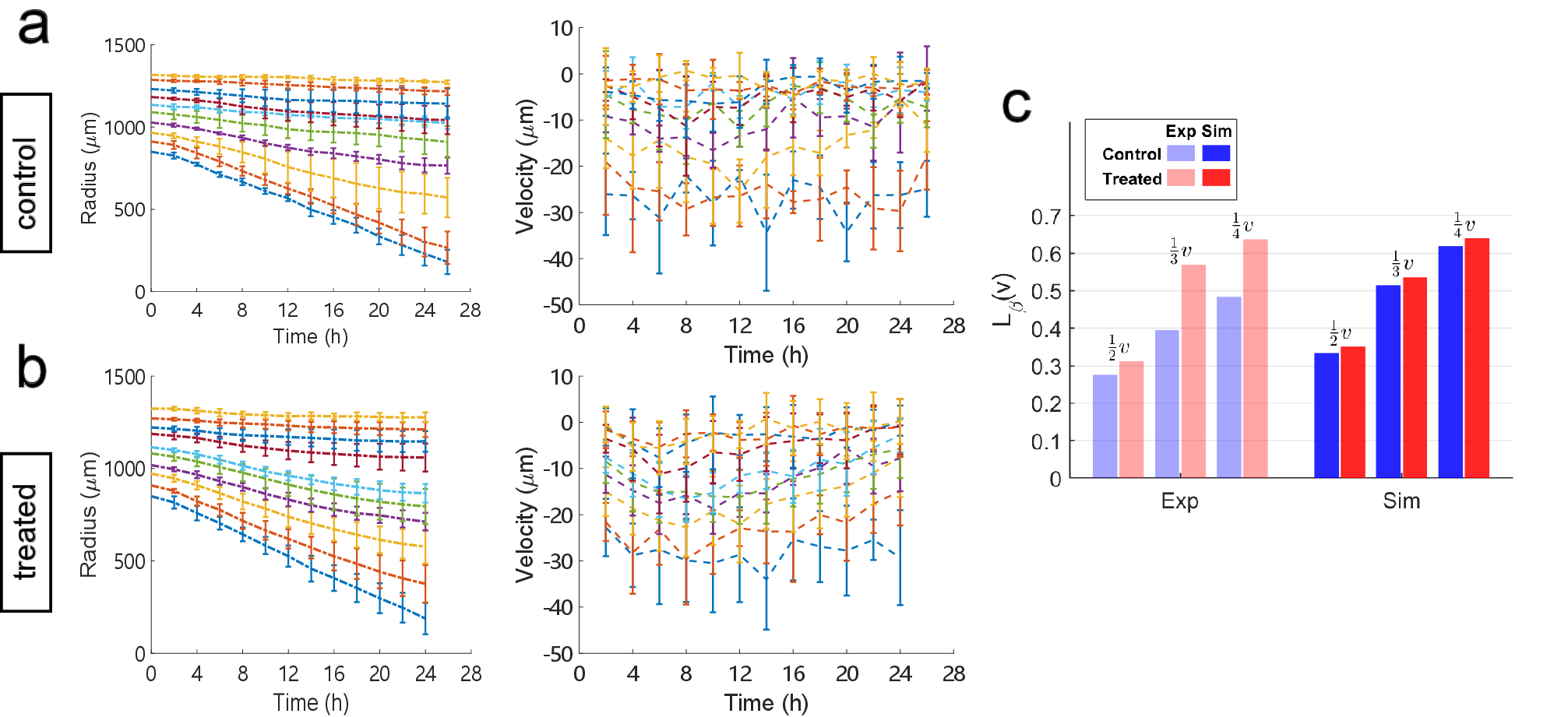}
\caption{Averaged radial position and inward radial velocity versus time for control (a) and treated (b) conditions, measured in experiments by tracking cells initially from 10 concentric layers with equal thickness. (c) The distance from the gap boundary to the locations where the radial velocity equals $1/2$, $1/3$, and $1/4$ of the boundary velocity, for comparison between experimental (Exp) and simulated (Sim) results, respectively.
}
\label{Fig1}
\end{figure}

\begin{figure}[htbp]
\centering
\includegraphics[width=\textwidth]{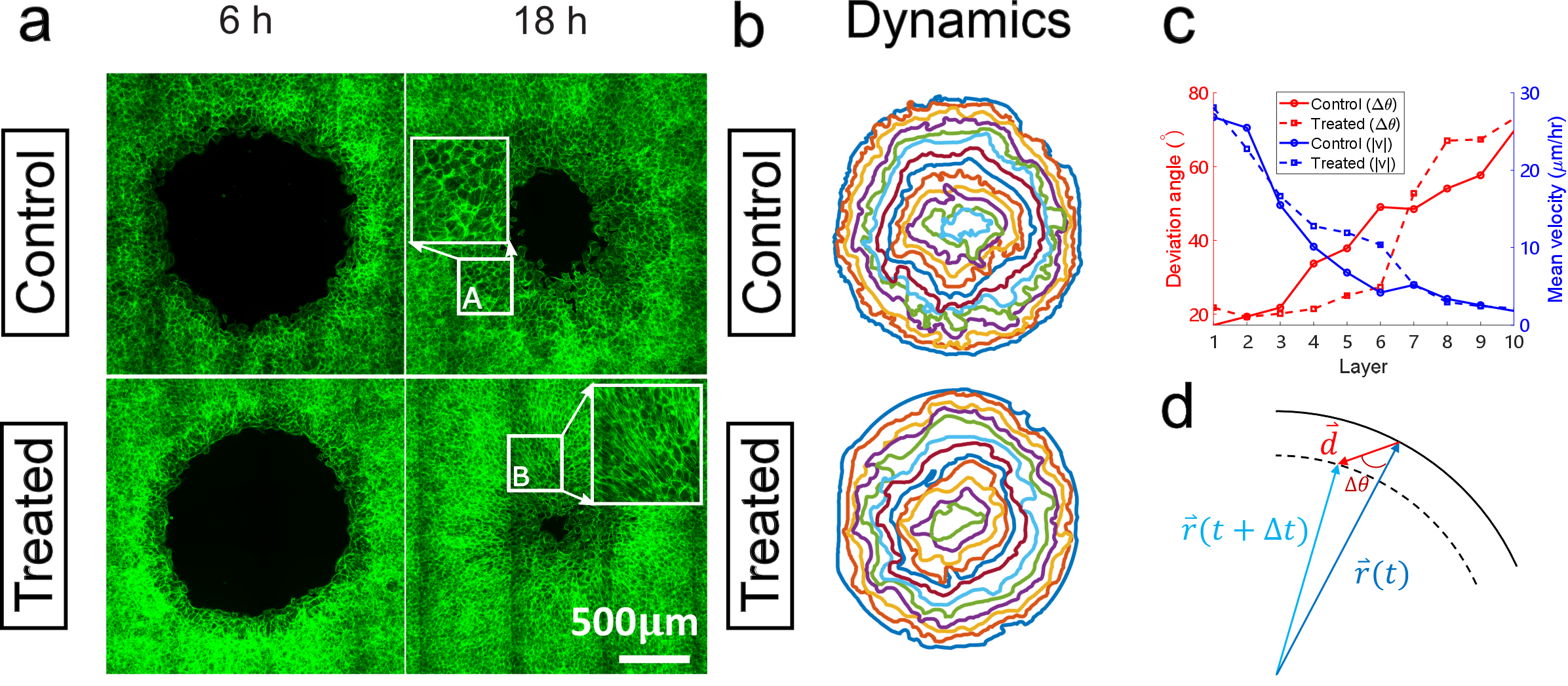}
\caption{(a) Tissue-level actin networks in MEC1 monolayers for control and treated conditions. Box A in control displays more dispersed actin fiber directions from the cell-cell junction directions, while box B in treated displays more radially aligned actin fibers. (b) The gap border dynamics during wound closure. (c) The mean velocity of 10 concentric layers and the mean deviation angle between the velocity and the radial direction, as illustrated in the diagram (d).
}
\label{Fig1S}
\end{figure}

\subsubsection{Inward tissue flow and deformation analysis}
The gap closures in the untreated (control) and treated conditions presented different dynamics. On average, in repeated experiments, the treated case reached closure around 3 hours faster than the control. Although both conditions exhibited quasi-circular wound closure (Fig.~\ref{Fig1S}b) (where local border undulations did not grow as in MDCK closures; \citealp{ben2014re}), we observed that the treated case maintained better border circularity than the control \citep{Jiangplaceholder2027placeholder}. To analyze the closing flow, we segmented the initial tissue outside of the circular gap into concentric layers of the same thickness (50 $\mu$m) from the gap front to the (almost immobile) outer regions. In particular, we tracked the averaged radial coordinates of the sampled cells in each layer, and computed their instantaneous inward velocities from the layer displacements over time (Fig.~\ref{Fig1}a,b). We quantified the layer-by-layer radial symmetry by computing the mean deviation angle between the instantaneous velocity and the radial direction (as shown in Fig.~\ref{Fig1S}d). Although both conditions showed quasi-circular inward flows, the treated case was indeed more radially symmetric than the control case in the front layers (L1--6) where the radial velocity was more significant (Fig.~\ref{Fig1S}c). We also noticed a growing speed gap between the first two layers and the following layers in the control case (Fig.~\ref{Fig1}a), which was not seen in the treated case (Fig.~\ref{Fig1}b). This suggests that the fast inward velocity in the control case was more localized near the gap boundary, whereas the treated case distributed inward motion over a broader region. This difference is further confirmed by quantifying the distances between the boundary and the locations where the velocity equals $1/2$, $1/3$, and $1/4$ of the boundary velocity (Fig.~\ref{Fig1}c, experiment).

\begin{figure}[htbp]
\centering
\includegraphics[width=\textwidth]{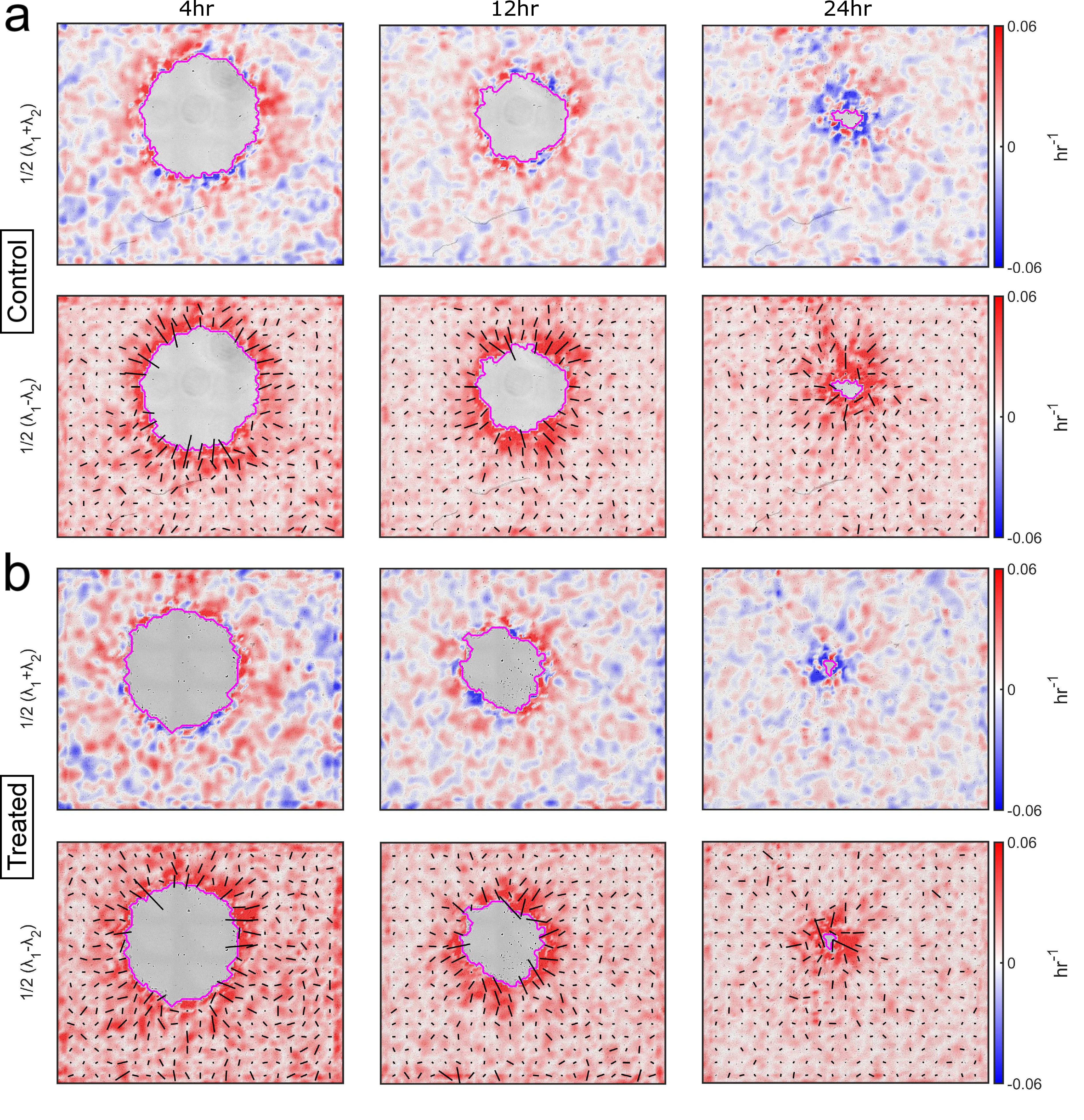}
\caption{The distribution of the isotropic strain-rate component $\tfrac{1}{2}(\lambda_1+\lambda_2)$ and the extension--contraction rate $\tfrac{1}{2}(\lambda_1-\lambda_2)$ in control (a) and treated (b) closing monolayers, where $\lambda_1$ and $\lambda_2$ are the principal values of $\mathbf{D}$. The color bars and the arrows indicate respectively the magnitude and direction of the strain rates.}
\label{Fig2}
\end{figure}

To analyze tissue deformation, we used Particle Image Velocimetry (PIV) to obtain the 2D velocity field $\mathbf{v}_{||}$, and computed the distributions of the planar strain rates $\mathbf{D}=\frac{1}{2}(\nabla_{\mathbf{x}}\mathbf{v}_{||}+\nabla_{\mathbf{x}}\mathbf{v}_{||}^{\mathsf T})$ \eqref{eq:strain_rate}. The isotropic component of $\mathbf{D}$ quantifies the rate of local area expansion or constriction $\tfrac{1}{2}\Tr(\mathbf{D})=\tfrac{1}{2}(\lambda_1+\lambda_2)$, while the deviatoric component $\mathbf{D}_D=\mathbf{D}-\tfrac{1}{2}\Tr(\mathbf{D})\mathbf{I}_2$ quantifies extension--contraction $\tfrac{1}{2}(\lambda_1-\lambda_2)$, where $\lambda_1$ and $\lambda_2$ are the principal values of $\mathbf{D}$. 
At the early and intermediate stages of closure, we observed significant areal growth in regions near, but not strictly localized to, the gap border (4 hr and 12 hr in Fig.~\ref{Fig2}). Such areal growth was partially due to global cell proliferation \citep{Jiangplaceholder2027placeholder}. At the late stage, we observed areal constriction localized around the gap border (24 hr in Fig.~\ref{Fig2}). This areal constriction appeared to differ from that seen in purse-string contraction in epithelial tissue \citep{Jiangplaceholder2027placeholder}, since we did not identify actin cables along the wound border throughout the closing process. Associated with the quasi-circular closure, the tissue underwent extension--contraction at all times, especially in regions closer to the wound (4 hr, 12 hr, and 24 hr in Fig.~\ref{Fig2}), where the direction of extension was mainly along the local normal direction to the gap border.

Together, these observations suggest that the main driving mechanism of closure is not a classical epithelial purse-string contraction. Instead, the inward tissue flow may arise from multiple coupled factors, including cell proliferation, elastic deformation, cell--cell intercalation, and traction with the substrate. Through systematic experimental analyses, we have ruled out traction with the substrate as a dominant driving force and identified cell proliferation as a key driver of the closing process in both cases \citep{Jiangplaceholder2027placeholder}. However, the complex flow and deformation patterns, including the different localization of inward motion, the radial extension--contraction pattern, and the radial cell elongation observed in the treated case, suggested different patterning of local elastic versus fluidic mechanical interactions. In the following, we apply our model to provide mechanical insight into the process and clarify the roles of growth, cell elastic deformation, and fluidic cell--cell intercalation in the control and treated cases.

\subsection{Planar fluidic rearrangements localize the inward flow}
We fit the model to the layer-resolved velocity data from Fig.~\ref{Fig1}a,b and obtained two calibrated parameter sets for the control and treated conditions; see Sec.~\ref{Sec.Fitting}, Fig.~\ref{Fig3S}, Table~\ref{tab:fiber-reinforcement} and Table~\ref{tab:no-fiber-reinforcement} for the fitting procedure and parameter values. Even though all model parameters were assumed to be spatiotemporally uniform within each condition, the calibrated simulations captured the qualitative trend of the inward tissue flow, including the difference among the two conditions. See Fig.~\ref{Fig1}c to compare experiments and simulations quantifying the locations where the radial velocity equals $1/2$, $1/3$, and $1/4$ of the boundary velocity in the control and treated conditions, respectively. The simulation qualitatively captured the same trend that the control case has more pronounced flow localization but with a slight underestimation of the difference, probably due to the uniform parameter assumption. 

\begin{figure}[htbp]
\centering
\includegraphics[width=\textwidth]{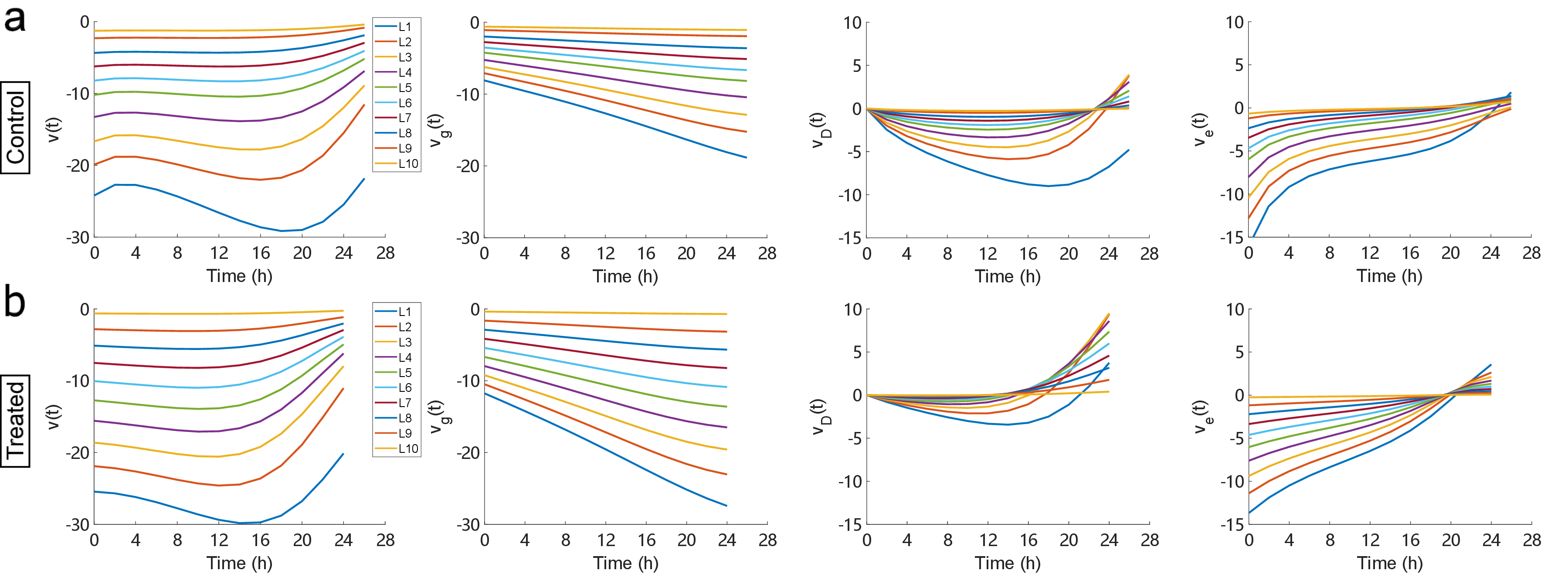}
\caption{Velocity of the 10 concentric layers from calibrated simulations for material points sharing the same initial coordinates from the experiments, and their decomposition into growth ($v_g$), elastic ($v_e$), and fluidic ($v_D$) sub-flows.}
\label{Fig3}
\end{figure}

We further compare the simulations with experiments by tracking simulated material points with the same initial positions as the experimental layers showed that the front layers moved faster than the trailing layers, and that the trailing layers slowed down significantly at late stages of closure (Fig.~\ref{Fig3}), similar to the experiments (see velocity plots in Fig.~\ref{Fig1}ab).

To identify how growth, elastic deformation, and fluidic rearrangement contribute to the inward flow, we decomposed the velocity gradient as $\nabla_{\mathbf{x}} \mathbf{v}_{||} = \boldsymbol{\Gamma}_{g_{||}}+ \boldsymbol{\Gamma}_{e_{||}} + \boldsymbol{\Gamma}_{D_{||}}$ (Fig.~\ref{Fig4S}), where the growth rate $\boldsymbol{\Gamma}_{g_{||}}=\mathrm{diag}(\gamma,\gamma)$ and the fluidic rearrangement rate $\boldsymbol{\Gamma}_{D_{||}} = \mathrm{diag}(\gamma_{Dr},\gamma_{D\theta})$ are directly given from \eqref{eq:radial_system}, and the elastic strain rates $\boldsymbol{\Gamma}_{e_{||}}=\mathrm{diag}(\gamma_{er},\gamma_{e\theta})$ can be computed thereby. 
Integrating the radial components of these three rates from the fixed outer boundary gives the corresponding sub-flows  $v_g=\int_{R_{out}}^{r}\gamma \,\mathrm{d}w$, $v_e=\int_{R_{out}}^{r}\gamma_{er} \,\mathrm{d}w$, and $v_D=\int_{R_{out}}^{r}\gamma_{Dr} \,\mathrm{d}w$, so that $v=v_g+v_e+v_D$ (Fig.~\ref{Fig3}). In both conditions, the growth sub-flow $v_g$ increased monotonically and provided a persistent inward contribution. By contrast, the elastic sub-flow $v_e$ decreased from the beginning and approached, or slightly crossed, zero near the end of closure, whereas the fluidic sub-flow $v_D$ first increased and then slowed down at later stages.

\begin{figure}[htbp]
\centering
\includegraphics[width=\textwidth]{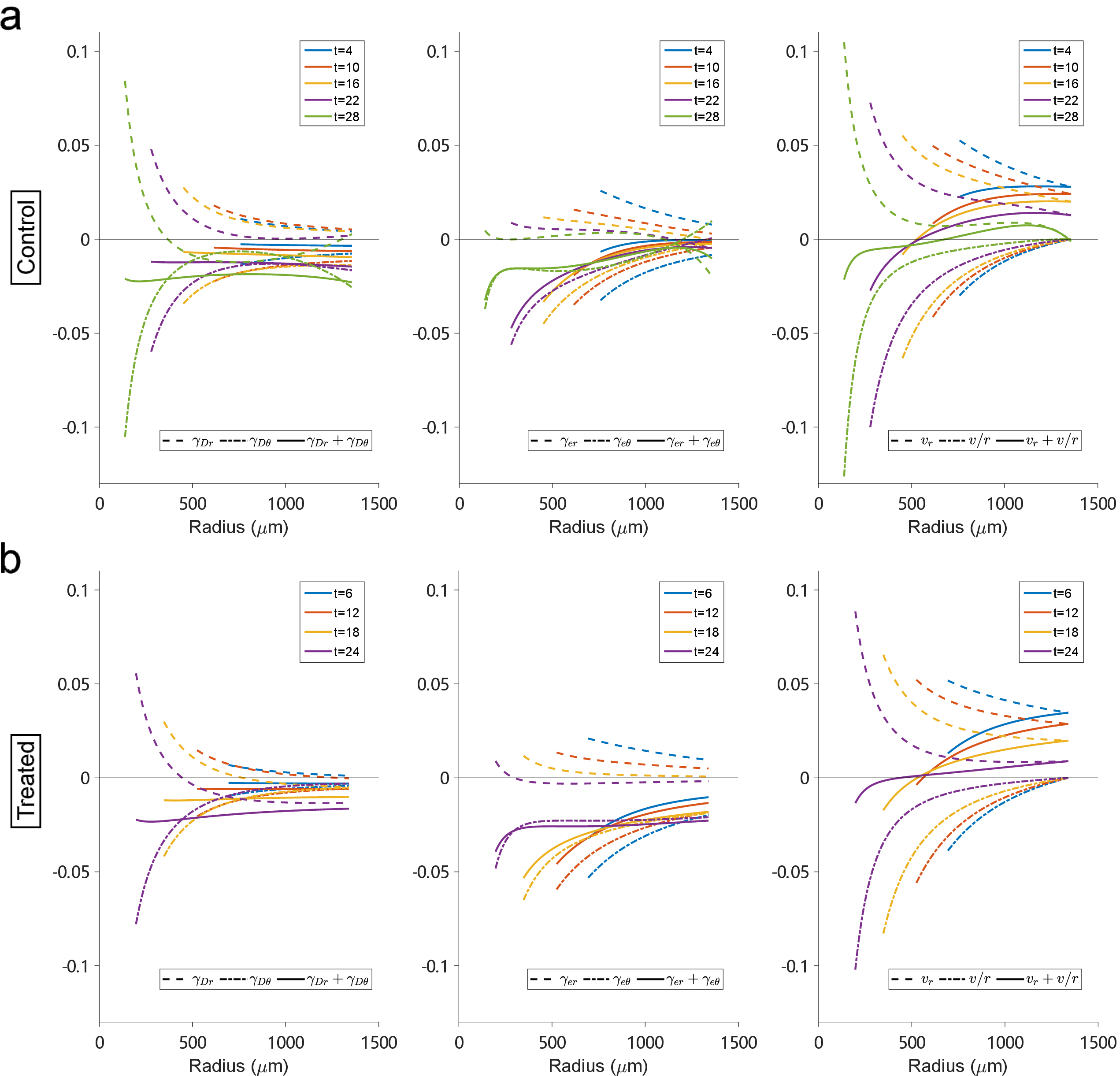}
\caption{The strain rates (radial, circumferential, and areal) (right) and the elastic (middle) and fluidic (left) contributions to the strain rates from calibrated simulations for control (a) and treated (b) cases with fiber-reinforcement.}
\label{Fig4}
\end{figure}

\begin{figure}[htbp]
\centering
\includegraphics[width=\textwidth]{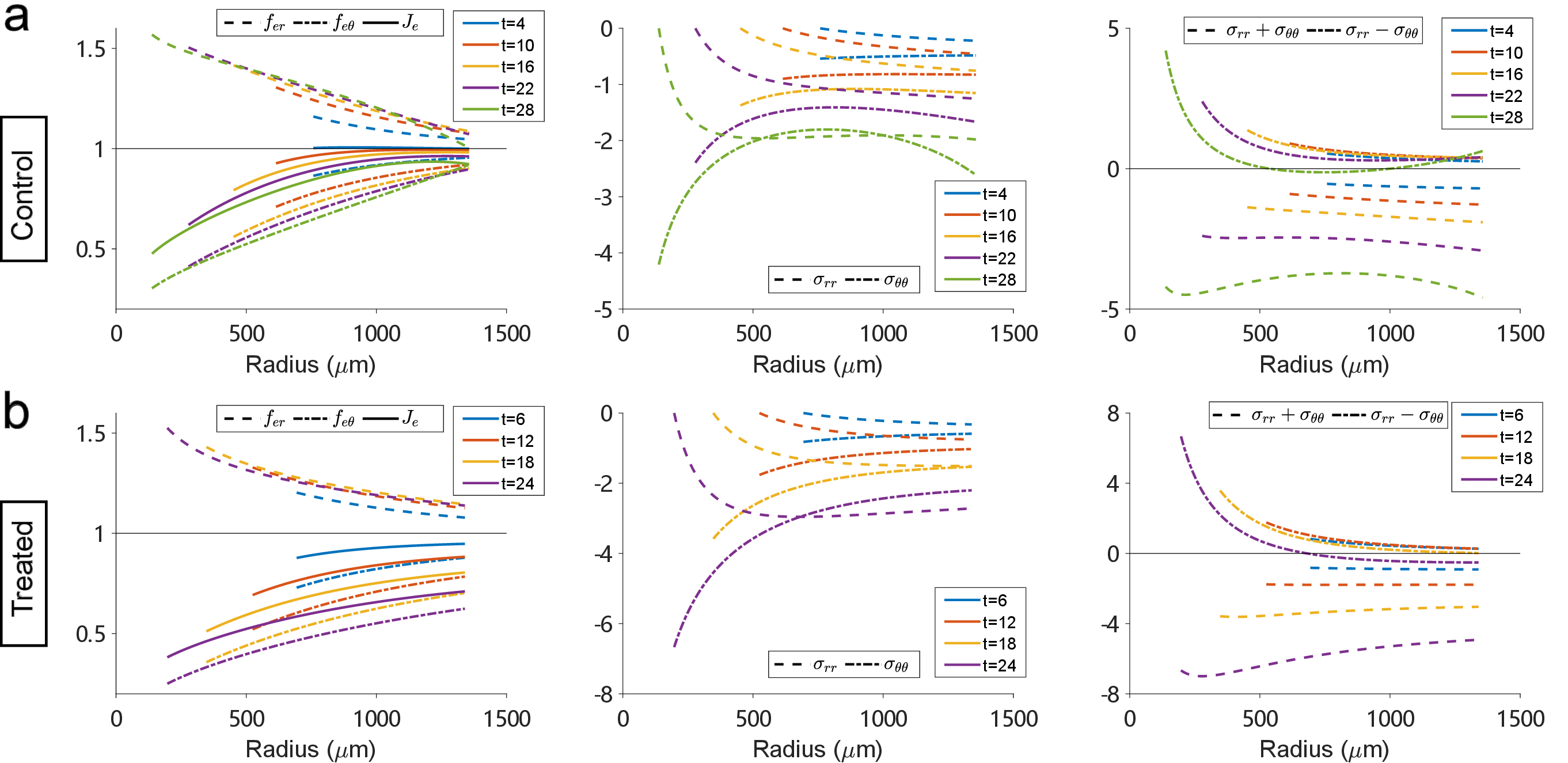}
\caption{The strains (radial, circumferential, and areal) (left), the stresses (middle), and their isotropic and deviatoric parts (right) from calibrated simulations for control (a) and treated (b) cases with fiber-reinforcement.}
\label{Fig4S}
\end{figure}

The decrease of $v_e$ can be interpreted through the radial elastic strain rate $\gamma_{er}$. As closure proceeds, $\gamma_{er}$ decreases and becomes negative in parts of the tissue at late stages (Fig.~\ref{Fig4}), indicating that radial elastic stretching is no longer being accumulated and is locally relaxing. Consistently, the accumulated stretch $f_{er}$ saturates or slightly decreases near the end of closure (Fig.~\ref{Fig4S}), and the radial integral of this decreasing or negative $\gamma_{er}$ leads to the slowdown or reversal of $v_e$. 

The turning point of $v_D$ is instead controlled by the evolving stress state. As tissue continues to move inward, $\sigma_{rr}$ becomes increasingly compressive and the rear region expands with larger radial compression than circumferential compression, as reflected by $\sigma_{rr}$ and $\sigma_{rr}-\sigma_{\theta\theta}$ in Fig.~\ref{Fig4S}. According to the radial rearrangement law \eqref{eq:radial_system}, both negative $\sigma_{rr}$ and negative $\sigma_{rr}-\sigma_{\theta\theta}$ promote radial contraction through fluidic rearrangement $\gamma_{Dr}$. This produces negative or weakened $\gamma_{Dr}$ in the rear region (Fig.~\ref{Fig4}), which in turn slows the fluidic sub-flow $v_D$.

The calibrated simulations also preserve the difference between the two conditions: the control case shows a more localized inward motion than the treated case, consistent with the experimental velocity-decay-length measurement in Fig.~\ref{Fig1}c. This difference is explained by the fitted planar fluidity. The control condition has a larger fluidity, with $\beta=0.03$, three times the treated value $\beta=0.01$ (Table~\ref{tab:fiber-reinforcement}). A larger $\beta$ amplifies the fluidic radial strain rate near the front (see $\gamma_{Dr}$ for two conditions in Fig.~\ref{Fig4}), producing a steeper spatial gradient in velocity (see $\tfrac{\partial v}{\partial r}$ for two conditions in Fig.~\ref{Fig4}) and thereby a more localized inward velocity profile. This model prediction that the control condition has higher fluidity is consistent with previous experimental evidence that the control condition has a higher cell-intercalation-associated strain rate than the treated condition \citep{Jiangplaceholder2027placeholder}.

\subsection{Fiber-reinforcement is key to local areal contraction}
Even though the fitted simulations assume a spatially uniform growth rate, the fiber-reinforced simulations reproduce the emergence of a front-localized areal constriction at late stages (as shown by the negative zone for the areal change rate $\partial_r v+v/r$ in Fig.~\ref{Fig4}), consistent with the experimental strain-rate maps in Fig.~\ref{Fig2}. The decomposition in Fig.~\ref{Fig4} further shows that the negative areal strain rate is due to 1) a global reduction of area from the third-dimensional rearrangement in response to the planar compression, $\gamma_{Dr}+\gamma_{D\theta}=\zeta(\sigma_{rr}+\sigma_{\theta\theta})/3$; and 2) the localized area reduction near the wound front, $\gamma_{er}+\gamma_{e\theta}$. Both constricting contributions enhance in time as the wound becomes smaller, which eventually brings $\partial_r v+v/r$  below zero near the wound front.

Thus, the constriction is a result of the elastic and fluidic response to the (uniform) growth and the geometry of the process, different from the purse-string mechanism. Here, the third-dimensional fluidic rearrangement is not associated with cell--cell intercalation, but rather a rearrangement of materials along the thickness of cell and cell--cell junction in response to the planar compression. Since the planar compression does not appear to be the largest at the wound border (see Fig.~\ref{Fig4S}), we hypothesize the constriction is mainly due to the elastic response through the elastic material property -- the fiber reinforcement. 

\begin{figure}[htbp]
\centering
\includegraphics[width=\textwidth]{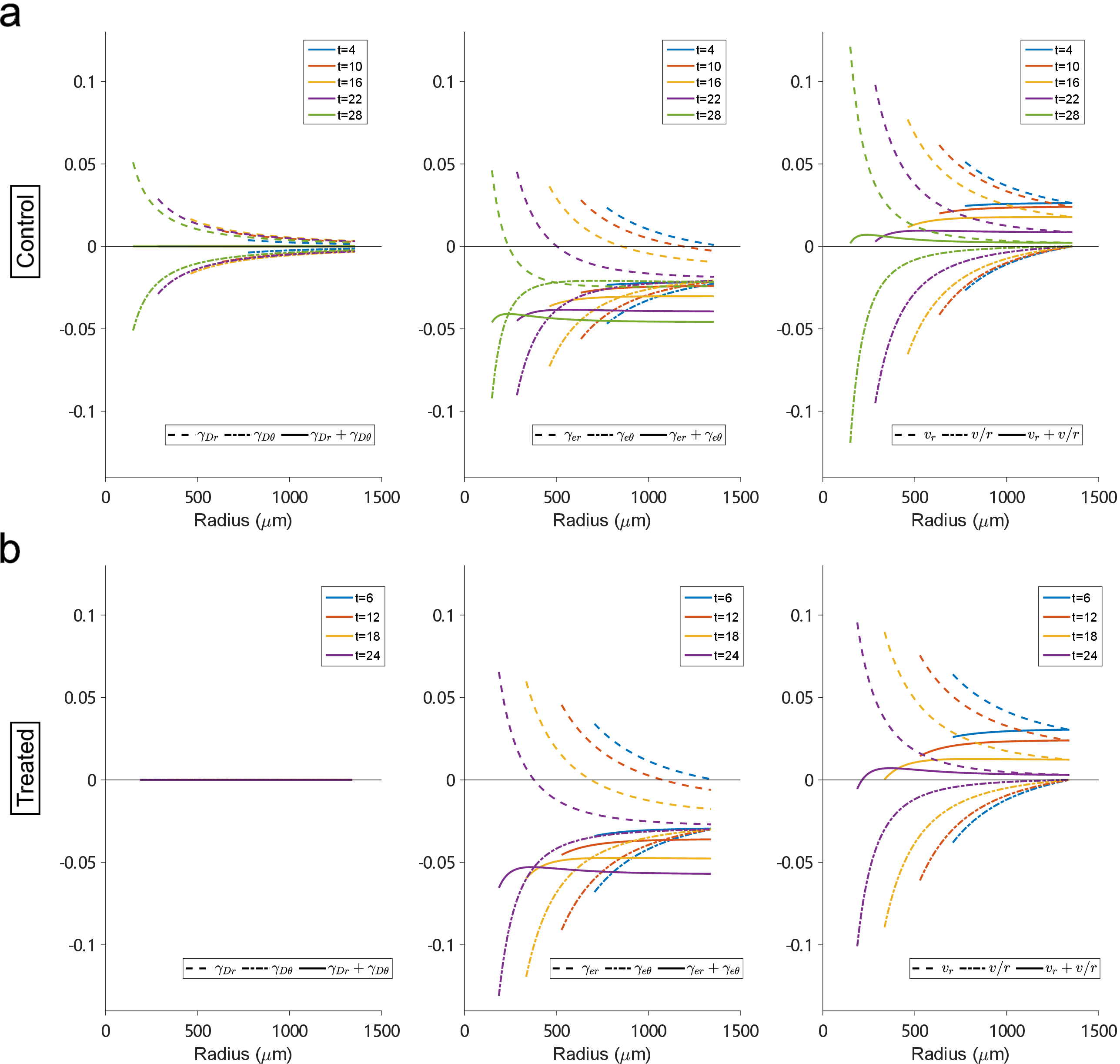}
\caption{The strain rates (radial, circumferential, and areal) (right) and the elastic (middle) and fluidic (left) contributions to the strain rates from calibrated simulations for control (a) and treated (b) cases without fiber reinforcement.}
\label{Fig5}
\end{figure}

To test this, we removed fiber reinforcement by setting $\eta=0$ and refit the velocity data from Fig.~\ref{Fig1}a,b with the remaining parameters. In the no-fiber refits, the control simulation no longer develops a clear constricting region, whereas the treated simulation shows only a small constriction zone and at a much later time than its fiber-reinforced counterpart (Fig.~\ref{Fig5}). 

\begin{figure}[htbp]
\centering
\includegraphics[width=\textwidth]{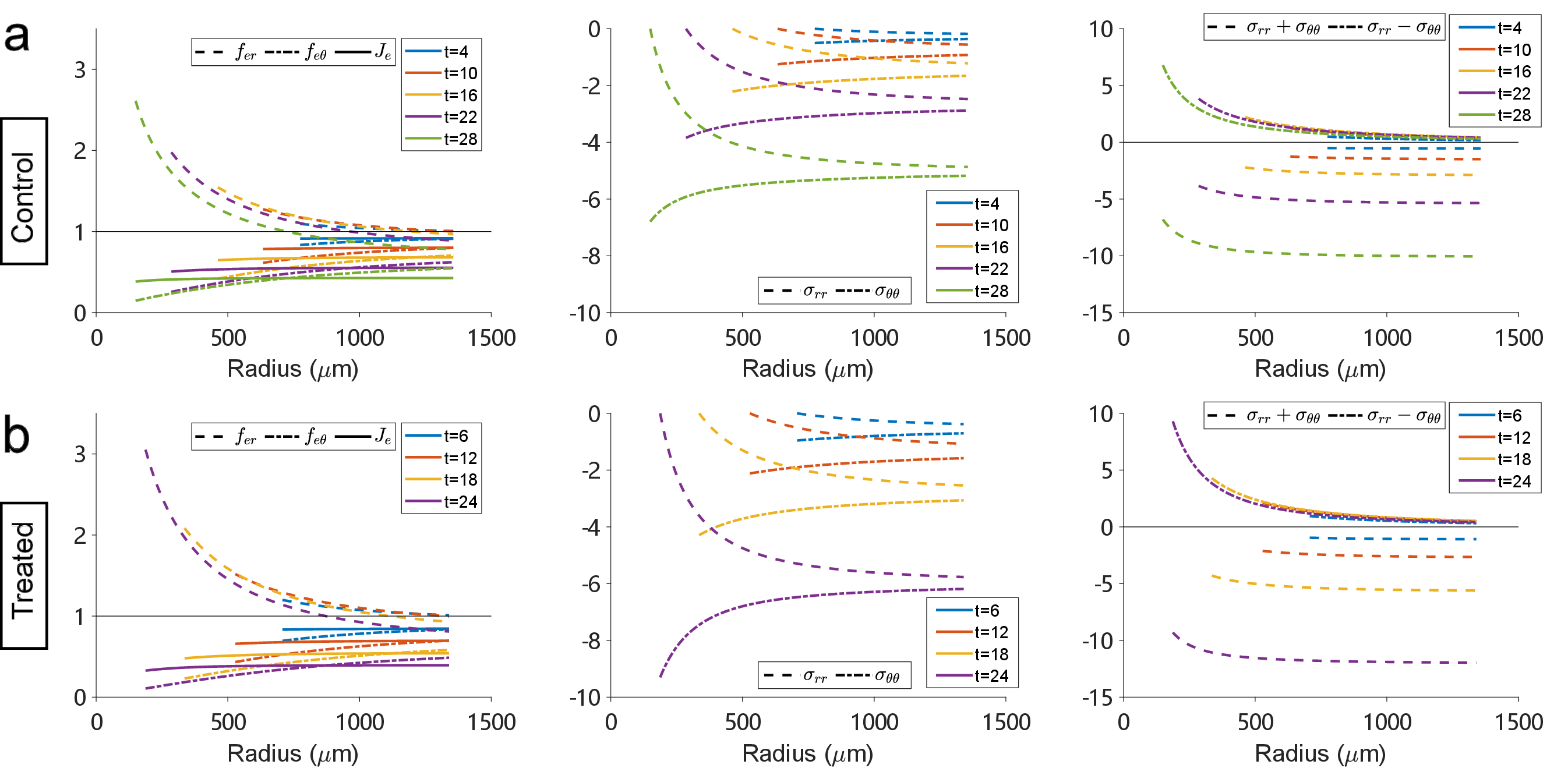}
\caption{The strains (radial, circumferential, and areal) (left), the stresses (middle), and their isotropic and deviatoric parts (right) from calibrated simulations for control (a) and treated (b) cases without fiber reinforcement.}
\label{FigS5}
\end{figure}

The comparison between fiber-reinforced and no-fiber simulations indicates that fiber reinforcement limits radial elastic accommodation. Without fiber reinforcement, the tissue develops a much larger radial elastic stretch $f_{er}$ (compare Fig.~\ref{FigS5} with Fig.~\ref{Fig4S}), increasing the capacity for radial extension. This radial extension can compensate for circumferential contraction near the wound front and thereby weaken the net areal constriction. With fiber reinforcement, radial extension is more limited, so circumferential elastic contraction is less compensated and appears as a negative areal strain rate in the front region. Thus, the fiber reinforcement contributes to the localized constriction by limiting the radial elastic deformation close to the boundary. 

Intriguingly, the no-fiber fits still preserve the relative difference in planar fluidity between conditions as with fiber-reinforcement: the control case retains a nonzero planar fluidity, $\beta=0.015$, while the treated case has $\beta=0$, indicating larger planar fluidity in the control case. For both conditions, the third-dimension fluidity is fitted to be $\zeta=0$ (Table~\ref{tab:no-fiber-reinforcement}).

\subsection{The synergistic roles of fiber-reinforcement and fluidity during closure}
\begin{figure}[htbp]
\centering
\includegraphics[width=\textwidth]{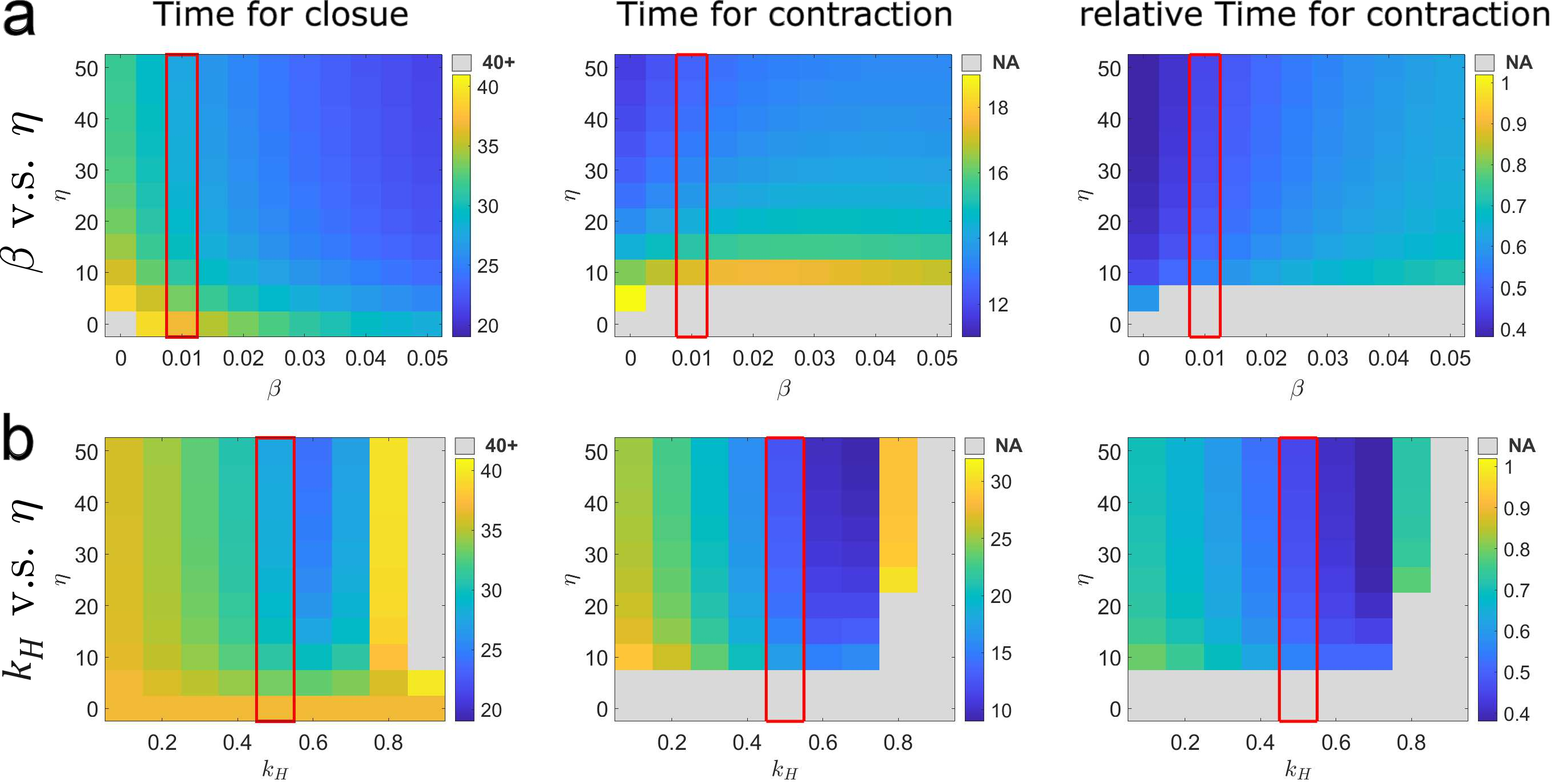}
\caption{Parameter maps showing the synergistic effects of fiber-reinforcement strength $\eta$, planar fluidity $\beta$, and fiber-alignment parameter $k_H$ on closure time, constriction-onset time, and the relative timing of constriction onset with respect to closure. The red frame is used to mark the same values of $\beta=0.01$ and $k_H=0.5$ used in row a) and b).}
\label{Fig6}
\end{figure}

Having identified planar fluidity as a regulator of inward-flow localization and fiber reinforcement as a key factor for front-localized areal constriction, we next asked how these mechanisms jointly determine closure dynamics. Since the growth rate $\gamma$ sets the overall time scale of the problem, we fixed $\gamma=0.02$ (the value between the control and treated cases) in this parameter study. We first varied the third-dimension fluidity $\zeta$ and found that increasing $\zeta$ substantially slowed closure, because the planar growth mainly contributes to thickness adjustment rather than in-plane expansion and inward motion. We therefore fixed $\zeta=0.015$ (same as the control case) at the reference value and focused on the coupled effects of planar fluidity $\beta$, fiber-reinforcement strength $\eta$, and fiber-alignment parameter $k_H$. For each simulation, we measured the time of wound closure, the time at which a negative areal strain-rate region first emerged, and the ratio between these two times (Fig.~\ref{Fig6}).

The parameter maps show that closure and front constriction are related but controlled by different mechanical ingredients. Increasing $\beta$ generally accelerated closure by strengthening stress-guided planar rearrangement and increasing the fluidic contribution to inward motion (Fig.~\ref{Fig6}a and Fig.~\ref{Fig7-beta}). However, stronger planar fluidity also relaxed elastic stretch and stress, so it did not by itself promote earlier elastic constriction. By contrast, increasing $\eta$ promoted the emergence of front-localized areal constriction and also shortened the closure time (Fig.~\ref{Fig6} and Fig.~\ref{Fig7-eta}). This agrees with the comparison in Sec.~3.3: fiber reinforcement limits radial elastic accommodation, and therefore yields the front-localized areal constriction. Thus, $\beta$ mainly controls how efficiently growth-induced stress is converted into inward flow, whereas $\eta$ controls whether the elastic state can generate the late-stage constricting region.

The dependence on $k_H$ was more subtle and non-monotonic, because changing $k_H$ redistributes fiber reinforcement between the radial and circumferential directions and thereby alters the stress anisotropy. Moderate alignment values produced the experimentally relevant pattern in which the constricting region remained localized near the wound front. In contrast, the $k_H=0.8$ case produced a qualitatively different mechanical state, where the areal constriction region occurred in the middle of the annulus (see $\partial_r v+v/r$ in Fig.~\ref{Fig7-kH}b), forming a ring-like constriction away from the wound front, which was not observed experimentally. In this regime, the strong circumferential bias in fiber reinforcement led to radial compression and retraction in the rear region, reflected by radial elastic compression ($f_{er}<1$) and by a pronounced valley of $\gamma_{er}$ in the middle region (Fig.~\ref{Fig7-kH}). This valley coincided with the transition position in $f_{er}$, indicating that the local radial retraction, rather than wound-front contraction, dominated the constricting pattern. Thus, although increasing $k_H$ can accelerate closure in part of the parameter space, the associated ring-like mechanical state does not represent the typical wound-front constriction observed during MEC1 monolayer closure.

\section{Discussion}\label{Discussion}
In this manuscript, we develop a nonlinear Maxwell-type growth-viscoelastic model to describe cell-monolayer tissue dynamics. We argue that this modeling approach offers a powerful new theoretical tool for analyzing tissue kinematics at the large-scale level and enables predictions of material properties and tissue fluidic remodeling that can be directly related to cellular and intercellular dynamics. The Maxwell-type viscosity is achieved by coupling the isochoric growth rate to the local deviatoric stress in an elastic-energy-dissipative manner within the framework of growth elasticity. While the multiplicative decomposition of the deformation gradient remains valid, we focus on tissue flow dynamics and explain the experimentally observed planar strain rate as the additive sum of the growth, elastic, and fluidic strain rates.

Through fitting the kinematic data with spatiotemporally uniform parameters, we recover the rates of both elastic and fluidic deformation throughout the course of closure.  In contrast to previous linear Maxwell fluid modeling \citep{dicko2017geometry,streichan2018global,ioratim2023mechanochemical}, our model describes the elastic deformation rate and reveals that it is quite significant in the fitted simulations. Specifically, the elastic deformation rate is the largest in the radial direction initially and gradually decreases as the elastic deformation saturates. 
Our results show that the elastic constitutive response is crucial to capturing the flow properties. In particular, fiber reinforcement is essential for reproducing the late-stage constriction during wound closure. Together, these findings highlight the importance of modeling elasticity in large-scale tissue flow dynamics. From our linear analysis (Sec.~\ref{subsec:Maxwell}), the proposed nonlinear elastic and fluidic model reduces to the linear Maxwell fluid model in the limit of small elastic strain. This analysis, together with the simulation results, implies that the previous linear Maxwell approaches are good approximations only when the fluidity is sufficiently large that the elastic strain rate remains small.

While the planar and vertical fluidities are kept constant, the geometry of the problem leads to different levels of stress as the contraction--elongation dynamics evolves. Specifically, the fluidic deformation rate gradually increases from zero and becomes significant later in the course of closure. Moreover, its contribution is most pronounced at the wound front, where the stress anisotropy is greatest. Between the two experimental conditions, the fitted simulations predict that one exhibits a higher level of fluidity, associated with a higher fluidic deformation rate. We show in \citet{Jiangplaceholder2027placeholder} that this prediction is qualitatively consistent with the contribution of cell--cell intercalation to tissue contraction--elongation observed in the experiments.

Besides the planar fluidity associated with cell--cell intercalation, we also model the vertical fluidity---the irreversible deformation along the thickness direction. Such deformation requires the vertical remodeling of adherens junctions, involving intercellular adhesion molecules (e.g., cadherins) and the subcellular cytoskeleton. While we do not directly model these cell-level and subcellular rearrangements, the model predicts thickness variation arising from such remodeling in response to the planar stress, in addition to the elastic deformation. Validation is left for future work.

\section*{Author contribution}
M. W. and C. W. built the mathematical model, interpreted the results and wrote the manuscript. H. J. performed experiments and collected data. C. W. designed and performed the simulations and analyzed the data. P. W. and Q. W. performed traction force microscopy analysis. N. O. and Y. G. performed PIV analysis. Funding was acquired by Y. S., M. W. and Q. W.  Y. S. and Q. W. provide feedback to the manuscript.

\section*{Acknowledgment}
This work is supported in part by the National Institute of Health (R35GM155279 and R01DK129990 to Y.S. and R01GM157590 to M. W. and Q. W.). We acknowledge the ADDFab and Light Microscopy Facilities at UMass Amherst for 3D printing and microscopy. M.W. acknowleges the partial support from the National Science Foundation – Division of Mathematical Sciences (NSF-DMS) under grants DMS-2012330 and DMS-2144372.

\appendix
\section{Numerical methods of reduced 2D mechanical model}\label{append:numerical}
By introducing the change of variable $\tilde{r}=\tfrac{r-R(t)}{R_{out}-R(t)}$ such that the moving boundary problem of the original system is reduced to a problem in a fixed domain for $\tilde{r}\in [0,1]$, we have the partial derivatives for any function $f(r,t)$
\begin{align}
&\partial_{r} f(r,t) = \partial_{\tilde{r}}f(\tilde{r},t)/\Delta R, \quad \Delta R = R_{out}-R(t), \\	
&\partial_{t} f(r,t) = \partial_{t}f(\tilde{r},t)-\frac{(1-\tilde{r})\dot{R}}{\Delta R}\partial_{\tilde{r}}f(\tilde{r},t)
\end{align}
From now on, we will drop the primes for convenience (i.e., $\tilde{r}\rightarrow r$). The system becomes
\begin{equation}
\begin{dcases}
&\frac{\partial {{f_e}_r}}{\partial t}+\tilde{v}\frac{\partial {{f_e}_r}}{\partial r}=\Big(\frac{1}{\Delta R}\frac{\partial v}{\partial r}-\gamma-\frac{\beta}{2}(\sigma_{rr}-\sigma_{\theta\theta})-\frac{\zeta}{3}(2\sigma_{rr}-\sigma_{\theta\theta})\Big){{f_e}_r}\\   
&\frac{\partial {{f_e}_\theta}}{\partial t}+\tilde{v}\frac{\partial {{f_e}_\theta}}{\partial r}=\Big(\frac{v}{R+r\Delta R} -\gamma-\frac{\beta}{2}(\sigma_{\theta\theta}-\sigma_{rr})-\frac{\zeta}{3}(2\sigma_{\theta\theta}-\sigma_{rr})\Big){{f_e}_\theta} \\
&\frac{1}{\Delta R}\frac{\partial \sigma_{rr}}{\partial r}+\frac{1}{R+r\Delta R} (\sigma_{rr}-\sigma_{\theta\theta})=0,\\
&\sigma_{rr}= \big(1+\eta(1-k_H)(I_k-1)\big){f_e}_r^2-{f_e}_r^{-2}{f_e}_\theta^{-2}, \\ 
&\sigma_{\theta\theta}= \big(1+\eta k_H(I_k-1)\big){f_e}_\theta^2-{f_e}_r^{-2}{f_e}_\theta^{-2},
\end{dcases}	
\end{equation}
where $\tilde{v}=(v-(1-r)\dot{R})/\Delta R$ and the initial and boundary conditions become
\begin{equation} 
\begin{dcases}
{f_e}_r(r,0)={f_e}_\theta(r,0)=1, \quad R(0)=R_0, \quad \text{at $t=0$}\\
\sigma_{rr}(0,t)=0,\quad \frac{dR}{dt} = v(0,t), \quad \text{at $r=0$}, \\
v(1,t)=0,  \quad \text{at $r=1$}\\
\end{dcases}
\end{equation}

We solve the above system by taking the time derivative of the mechanical equilibrium equation and converting the coupled system to a nonlinear system of the velocity only, and then applying a semi-implicit finite difference scheme to numerically solve velocity fields. The other variables such as the strains and stresses are then recovered from the velocity field. We refer the readers to \citet{olaranont2025chemomechanical} for the details of the numerical methods. 

\section{Parameter fitting}
\label{Sec.Fitting}
\begin{figure}[htbp]
\centering
\includegraphics[width=0.7\textwidth]{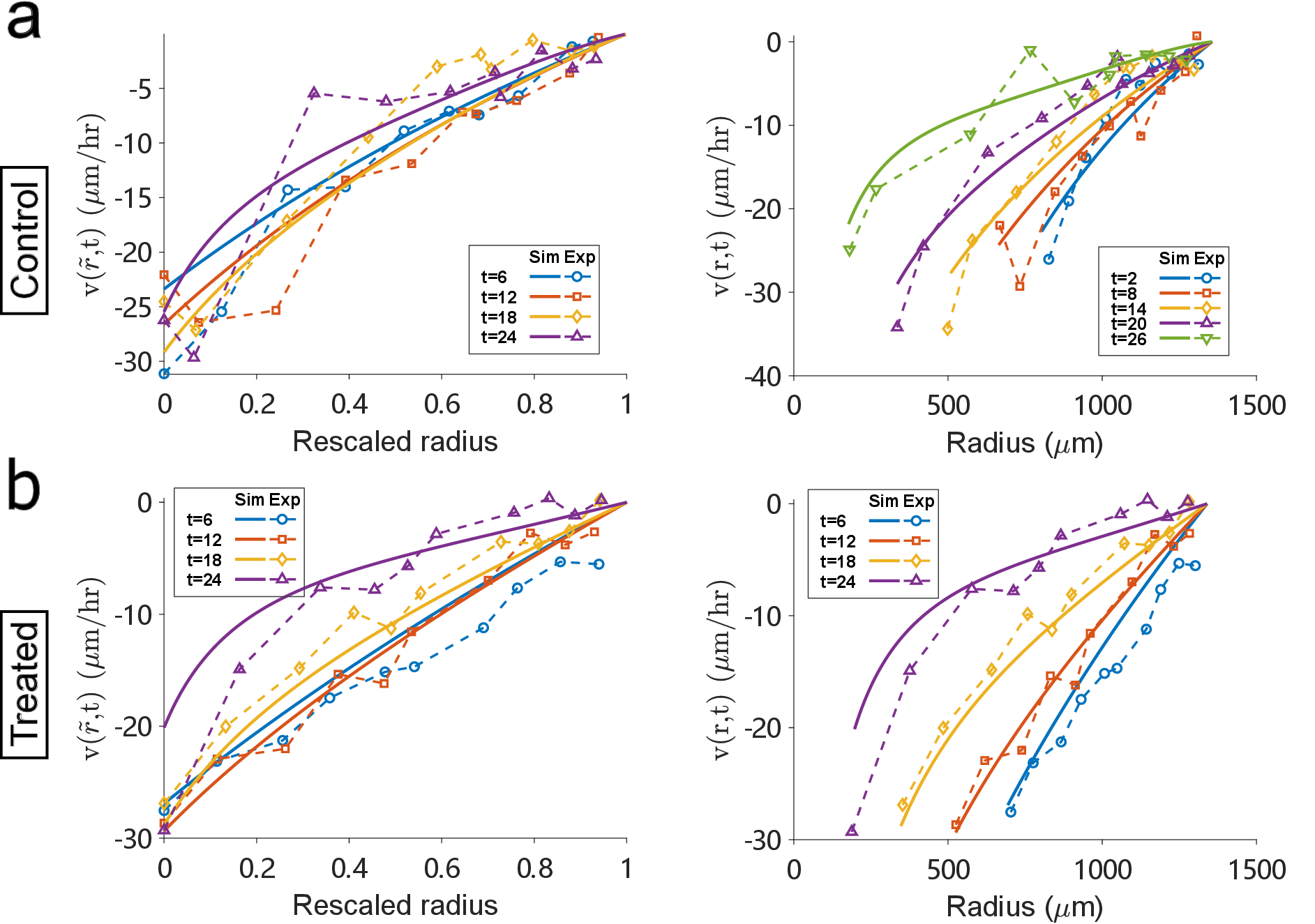}
\caption{Velocity fitting in the rescaled coordinates (left) and in the Eulerian coordinates (right) for the control (a) and treated (b) cases. The rescaled coordinates are defined as $\tilde{r}:=r/R(t)$, where $R(t)$ is the gap radius at time $t$.}
\label{Fig3S}
\end{figure}

We fit the model to the experimental data by minimizing the error between the simulated and measured inward velocities of the 10 concentric layers. The error is defined as $\sqrt{\sum_i \left(v_i^k - v_i(t_k)\right)^2}$, where $v_i^k$ is the measured velocity of layer $i$ at time $t_k$, and $v_i(t_k)$ is the corresponding simulated velocity. To obtain the simulated velocities $v_i(t_k)$ for each layer in Fig.~\ref{Fig3S}, we track the material points that share the same initial coordinates as the experimental layers, and obtain their inward velocities by interpolating the simulated (Eulerian) velocity field $v(r)$ at their current positions at each time point. The current positions of the material points are obtained by the inverse of the reference map introduced by \citet{Wei2023elasticmodel} ($y(r)$, i.e., the mapping from the current configuration $r$ back to the reference configuration $y$), which is updated at each time step by integrating the simulated velocity field via $\partial_t y(r)= -v(r) \partial_r y(r)$.

We performed a grid search over the parameter space for $\beta$, $\zeta$, $\eta$, $k_H$, and $\gamma$ to find the best-fit parameters that minimize this error. The best-fit parameters with and without fiber reinforcement for both control and treated conditions are summarized in Table~\ref{tab:fiber-reinforcement} and Table~\ref{tab:no-fiber-reinforcement}, respectively.

\begin{table}[htbp]
\centering
\caption{Best-fit parameter combinations with fiber reinforcement.}
\label{tab:fiber-reinforcement}
\renewcommand{\arraystretch}{1.6}
\resizebox{\textwidth}{!}{%
\begin{tabular}{|l|c|c|c|c|c|c|}
\hline
\begin{tabular}[c]{@{}l@{}}With fiber-\\reinforcement\end{tabular} &
\begin{tabular}[c]{@{}c@{}}$0 \le \beta \le 0.05,$\\$\Delta\beta = 0.005$\end{tabular} &
\begin{tabular}[c]{@{}c@{}}$0 \le \zeta \le 0.05$\\$\Delta\zeta = 0.005$\end{tabular} &
\begin{tabular}[c]{@{}c@{}}$10 \le \eta \le 100$\\$\Delta\eta = 10$\end{tabular} &
\begin{tabular}[c]{@{}c@{}}$0.3 \le k_H \le 0.7$\\$\Delta k_H = 0.1$\end{tabular} &
\begin{tabular}[c]{@{}c@{}}Control: $0.012 \le \gamma \le 0.024$\\Treated: $0.02 \le \gamma \le 0.028$\\$\Delta\gamma = 0.002$\end{tabular} &
\begin{tabular}[c]{@{}c@{}}Error:\\$\displaystyle \sqrt{\sum_i \left(v_i^k - v_i(t_k)\right)^2}$\end{tabular} \\
\hline
Control & $0.03$ & $0.015$ & $50$ & $0.6$ & $0.016$ & $35.4737$ \\
\hline
Treated & $0.01$ & $0.01$ & $20$ & $0.5$ & $0.024$ & $25.4012$ \\
\hline
\end{tabular}%
}
\end{table}

\begin{table}[htbp]
\centering
\caption{Best-fit parameter combinations without fiber reinforcement.}
\label{tab:no-fiber-reinforcement}
\renewcommand{\arraystretch}{1.6}
\resizebox{\textwidth}{!}{%
\begin{tabular}{|l|c|c|c|c|}
\hline
\textbf{No fiber-reinforcement} &
\begin{tabular}[c]{@{}c@{}}$0 \le \beta \le 0.05,$\\$\Delta\beta = 0.005$\end{tabular} &
\begin{tabular}[c]{@{}c@{}}$0 \le \zeta \le 0.02$\\$\Delta\zeta = 0.005$\end{tabular} &
\begin{tabular}[c]{@{}c@{}}Control: $0.022 \le \gamma \le 0.03$\\Treated: $0.024 \le \gamma \le 0.036$\\$\Delta\gamma = 0.002$\end{tabular} &
\begin{tabular}[c]{@{}c@{}}Error:\\$\displaystyle \sqrt{\sum_i \left(v_i^k - v_i(t_k)\right)^2}$\end{tabular} \\
\hline
Control & $0.015$ & $0$ & $0.024$ & $37.3829$ \\
\hline
Treated & $0$ & $0$ & $0.03$ & $29.6659$ \\
\hline
\end{tabular}%
}
\end{table}

\FloatBarrier
\section{Supplementary figures for parameter study}

\begin{figure}[H]
\centering
\includegraphics[width=0.95\textwidth]{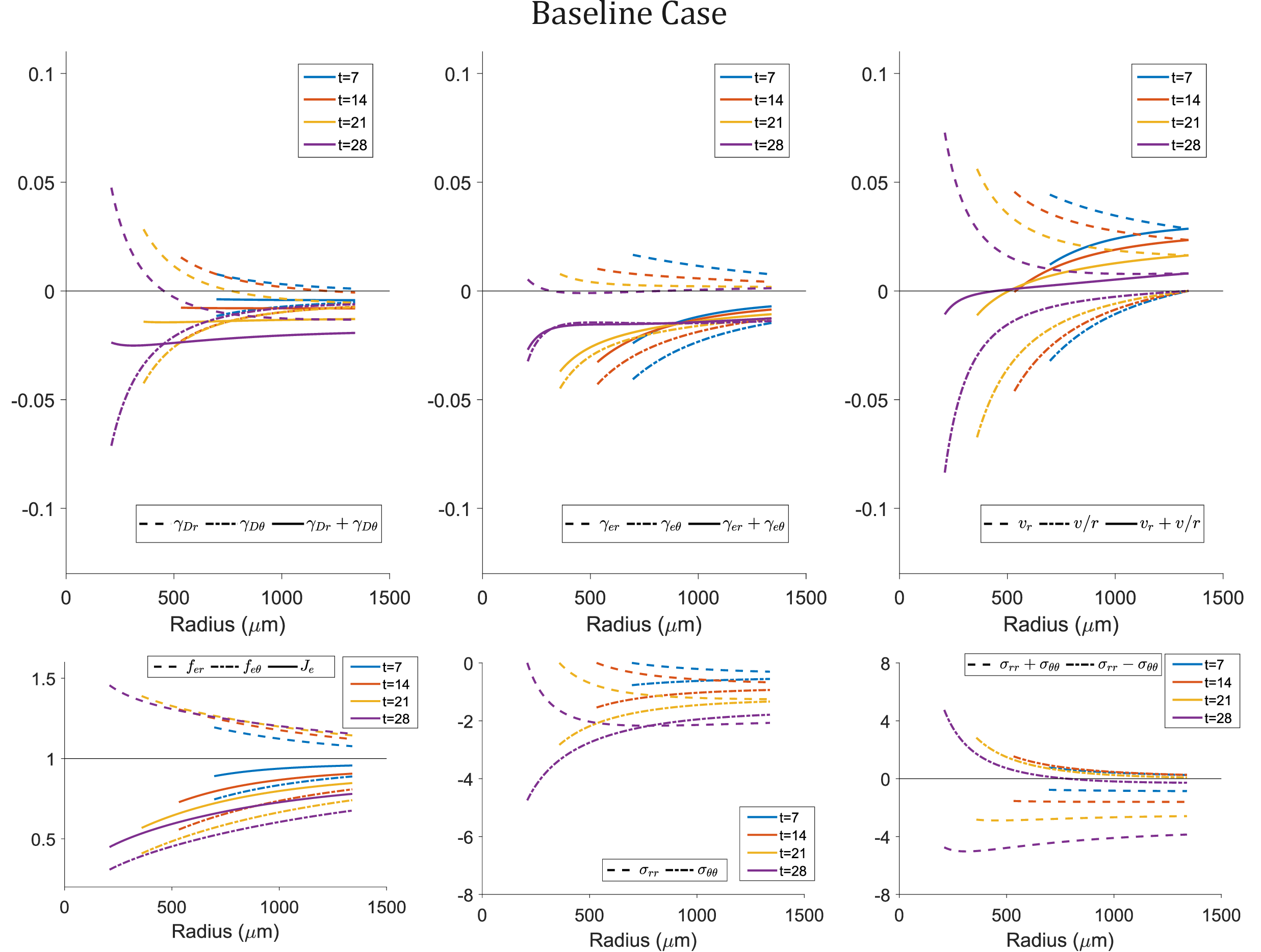}
\caption{Strain-rate decomposition (top) and mechanical state variables (bottom) for the reference case with $\beta=0.01$, $\zeta=0.015$, $\eta=25$, $k_H=0.5$, $\mu=1$, and $\gamma=0.02$.}
\label{Fig7-ref}
\end{figure}

\begin{figure}[htbp]
\centering
\includegraphics[height=0.975\textheight,keepaspectratio]{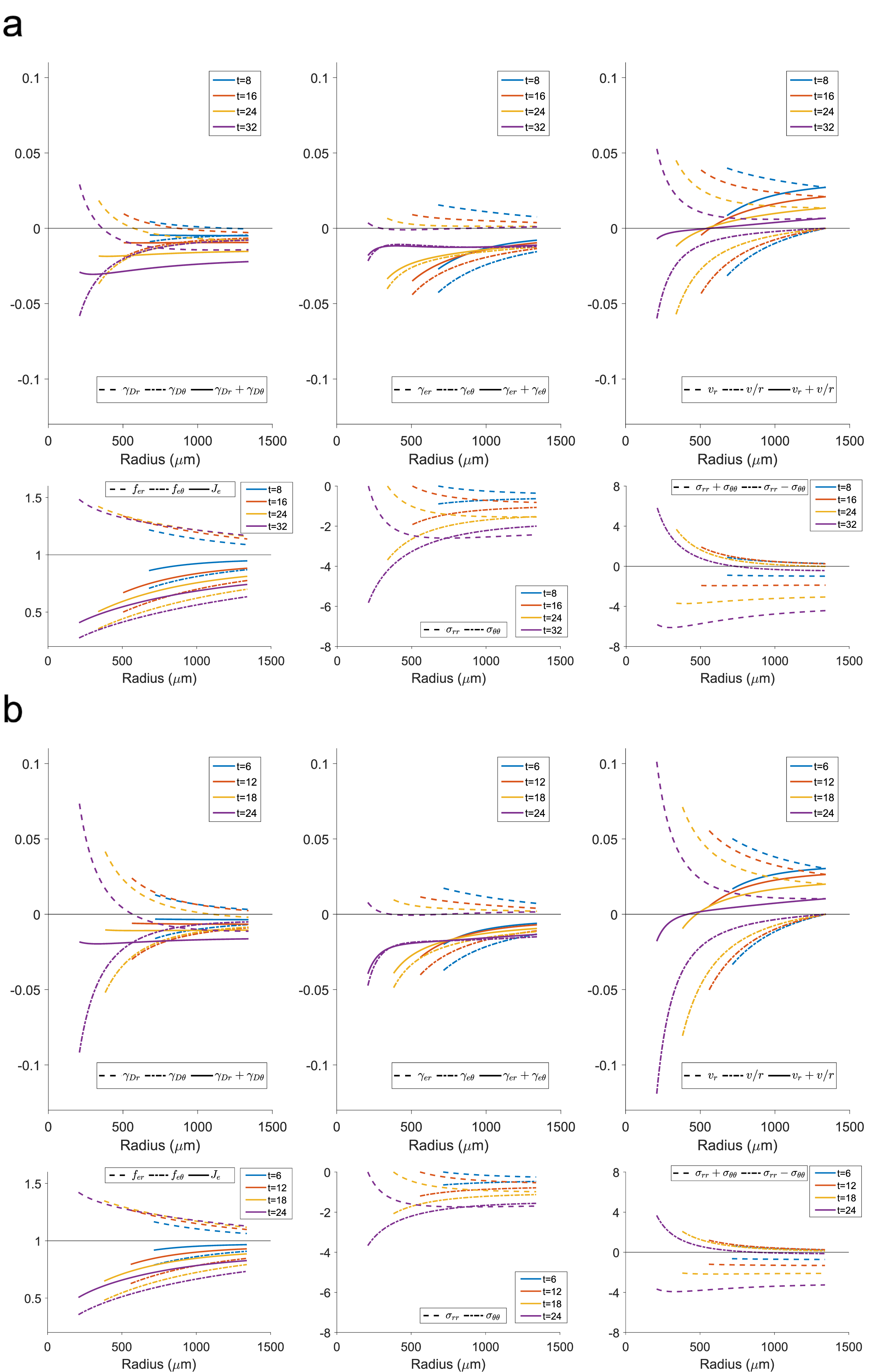}
\caption{Effect of planar fluidity on strain-rate decomposition (top) and mechanical state variables (bottom), comparing $\beta=0$ (a) and $\beta=0.03$ (b).}
\label{Fig7-beta}
\end{figure}

\begin{figure}[htbp]
\centering
\includegraphics[height=0.975\textheight,keepaspectratio]{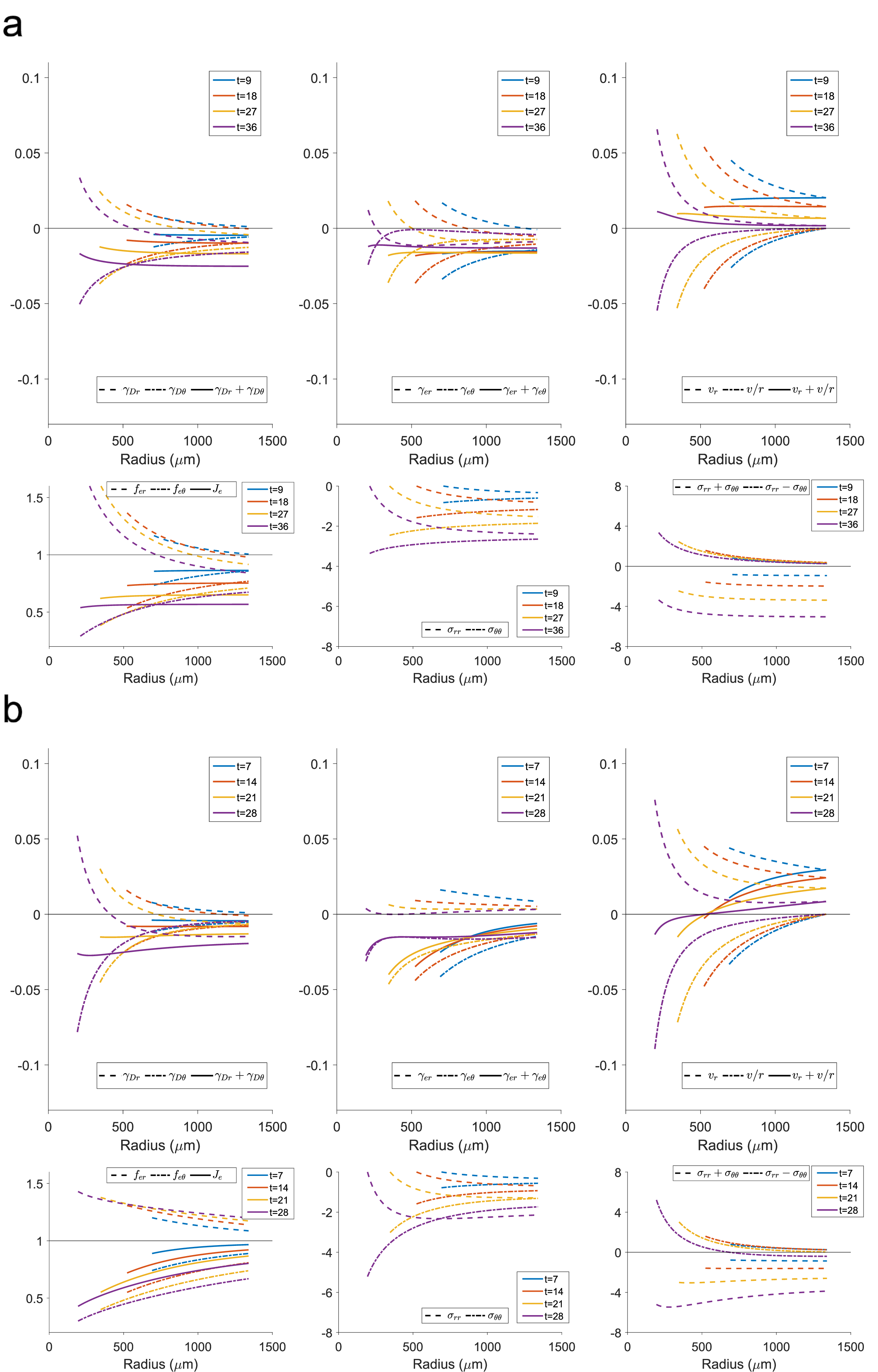}
\caption{Effect of fiber-reinforcement strength on strain-rate decomposition (top) and mechanical state variables (bottom), comparing $\eta=0$ (a) and $\eta=50$ (b).}
\label{Fig7-eta}
\end{figure}

\begin{figure}[htbp]
\centering
\includegraphics[height=0.975\textheight,keepaspectratio]{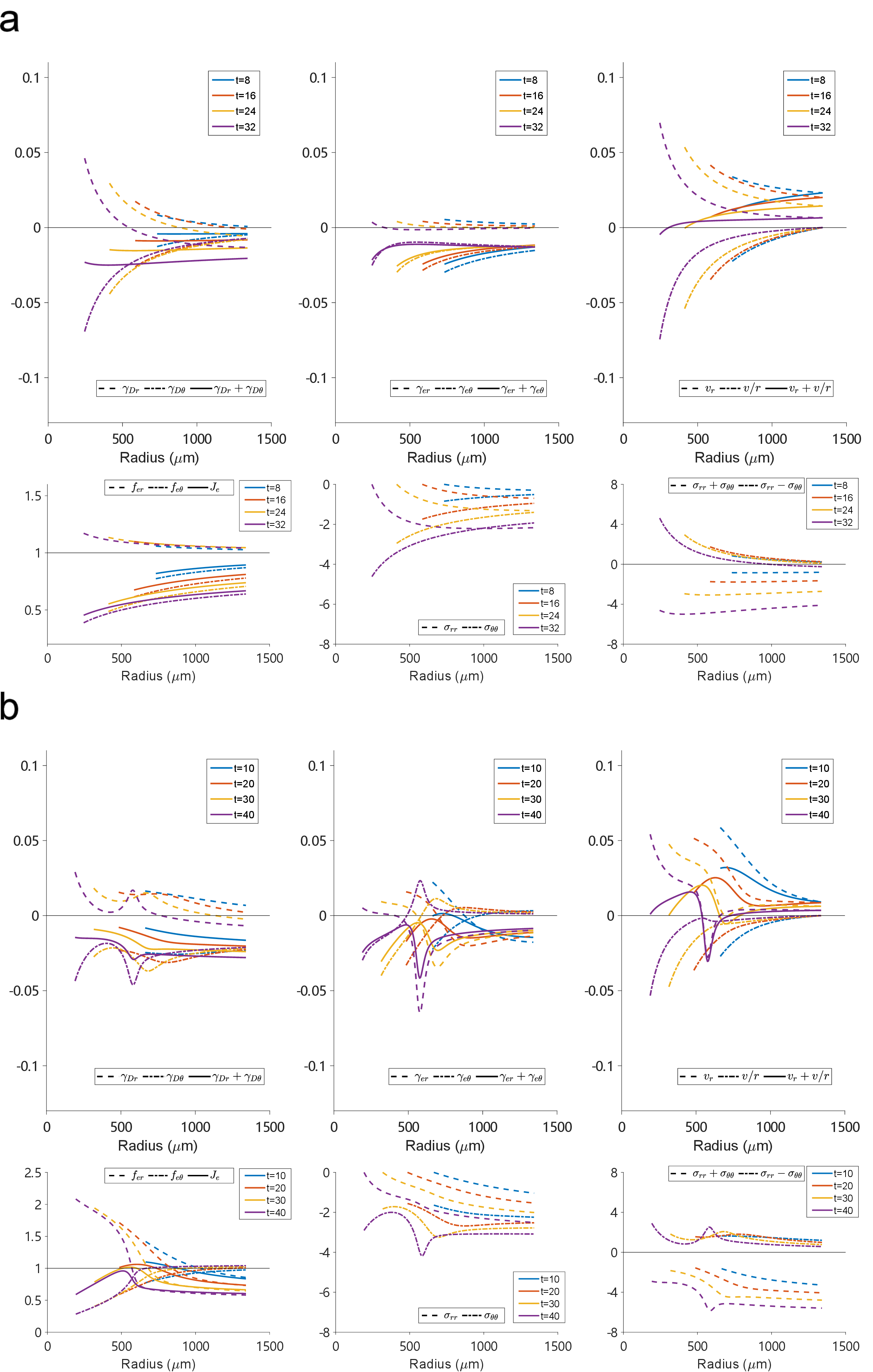}
\caption{Effect of fiber-alignment parameter on strain-rate decomposition (top) and mechanical state variables (bottom), comparing $k_H=0.2$ (a) and $k_H=0.8$ (b). }
\label{Fig7-kH}
\end{figure}

\FloatBarrier
\bibliographystyle{elsarticle-harv}
\bibliography{refs}

\end{document}